# Effects of Solar Eclipse of March 20, 2015 on the Ionosphere

*Dario Sabbagh*[(1,2)], *Carlo Scotto*[(1)], *Alessandro Ippolito*[(1)], *Alessandro Settimi*[(1)], *Vittorio Sgrigna*[(2)]

(1) Istituto Nazionale di Geofisica e Vulcanologia,
Via di Vigna Murata 605, I-00143 Roma, Italy

(2) Università degli Studi Roma Tre, Dipartimento di Matematica e Fisica,
Via della Vasca Navale 84, I-00146 Roma, Italy

**Abstract**

The effects of the solar eclipse of March 20, 2015 on different ionospheric layers were studied, using vertical ionospheric soundings from the ionosondes of Rome (41.8N, 12.5E), Gibilmanna (37.9N, 14.0E), and San Vito dei Normanni (40.6N, 18.0E). The responses of the critical frequencies were investigated during the solar eclipse, and the formulations used for their estimation were corrected taking into account the decreased solar irradiance. This effect was modeled as a Solar Obscuration Factor (*SOF*) and comparisons with experimental values were performed. A further study on the occurrence of the sporadic E layer during the eclipse is presented. When ionogram analysis is limited to 3 days before and 3 days after the eclipse, the appearance of the sporadic E layer seems to be related to the eclipse. However, when a longer range of days before and after the eclipse event are taken into account this phenomenon does not appear so clear. The behavior of a regional adaptive and assimilative 3D ionospheric model was also tested, assimilating plasma frequency profiles $f_p(h)$. Studying the behavior of the model in such unusual conditions enabled the introduction of corrections to the $f_oE$, $f_oF_1$, and $f_oF_2$ formulations, improving the performances of the model itself.

## 1. Introduction

During a solar eclipse it is possible to investigate changes in the ionosphere as a response to the sudden decrease in solar radiation. Since the first half of the twentieth century, it was assumed that an eclipse would affect the ionization of the upper and lower layers of the ionosphere, therefore radio-telegraphic investigations were performed during the solar eclipse involving South America and Africa on May 29, 1919 (Eccles, 1919). A reduction in plasma density was observed in some subsequent papers (Gledhill, 1959a; Gledhill, 1959b) and attempts were made to establish the recombination coefficients (Beynon,1955). Irregularities, tilts, and sporadic E ($E_s$) layers were also observed and reported by numerous researchers (McLeish, 1948; Minnis, 1954; Bramley, 1956; Minnis, 1956; Bates and McDowell, 1957). An interesting irregularity which received attention during early observations of the eclipse, was the temporary appearance of a critical frequency between the F1 and F2 layers (Beynon, 1955).

As regards the evaluation of time constants, there are now consolidated studies in the literature that take into account variations in electron density, temperature, and diffusion processes (Baron and Hunsuker, 1973; Cohen, 1984; Mueller-Wodarg et al., 1998; Huang et al., 1999; McPherson et al., 2000).

The recent and comprehensive achievements reported by different authors for the E, F1, and F2 mid-latitude regions are very interesting in this context. A lot of information was obtained by studying the ionospheric responses to the eclipse that occurred on August 11, 1999, with a path of totality passing over central Europe so that it could be measured by an extensive ionosonde network (Stanislawska et al. 2001; Le et al. 1999; Farges et al. 2001).

As a result, today it is considered proven that the redistribution of plasma during an eclipse can lead to changes in the form of $N_e(h)$, observable using ionosondes. It is also considered certain that the relative decrease in the critical frequencies is most noticeable in the lower ionosphere, where recombination is efficient. If dynamic forces are ignored, such as thermospheric winds or electric fields, an increase should be observed in the peak height $h_mF_2$ as well as an increase of the layer thickness expressible through $B_0$. In quiet conditions, it is reasonable to ignore dynamic forces like thermospheric winds or electric fields, and diffusive transport processes are expected to dominate. In these circumstances a delayed response to obscuration should be observed, and the delay should increase with height (Jackowski et al., 2008).

The relationship between eclipse events and $E_s$ layers has also been studied during several eclipses. Some authors relate $E_s$ occurrence to internal gravity waves (GWs) generated in the atmosphere by the fast-moving cooling spot of the shadow region (Datta, 1972; Datta, 1973; Chen et al., 2010; Yadav et al., 2013; Pezzopane et al., 2015). This explanation is in line with the theory that suggests atmospheric GWs play a significant role in the wind-shear mechanism related to $E_s$ layer formation at mid-latitudes (Axford, 1963; Chimonas and Axford, 1968). Nevertheless, with the introduction of height-time-intensity (HTI) analysis for $E_s$ investigation, Haldoupis et al. (2006) suggested that $E_s$ occurrence and periodicity are mainly related to the downward propagation of wind-shear convergent nodes associated with diurnal and semidiurnal tides in the E region, even though GWs can be important at times for reinforcing or disrupting the tidal forcing.

In this work we investigate the solar eclipse-induced decrease of $f_oE$, $f_oF_1$, $f_oF_2$ and the possible relationship between eclipse events and $E_s$ occurrence. This task is accomplished by studying the effects of the March 20, 2015 partial solar eclipse on various ionospheric layers, using the measurements of ionosondes operating in Italy. We also try to establish the relationships between the Solar Obscuration Factor (*SOF*) and critical frequencies in order to evaluate whether these relationships can be usefully introduced in order to simulate the effects of an eclipse in a regional adaptive and assimilative 3D ionospheric model.

## 2. Characterization of an eclipse with the Solar Obscuration Factor (*SOF*)

This work applied a fairly simple approach for modeling the time behavior of *SOF* during the evolution of an eclipse, with *SOF* defined as the fraction of the Sun's area blocked by the Moon and expressed as a percentage. The applied model is similar to that used by Möllmann and Vollmer (2006), with the Sun and Moon assumed to follow a straight and uniform motion, and both to be circular in shape, ignoring any effects of limb darkening. The input parameters include the beginning of the eclipse, its duration, and the value of *SOF* at maximum. Fig. 1 shows the time behavior of *SOF* obtained at the ionospheric stations of Rome (41.8N, 12.5E), Gibilmanna (37.9N, 14.0E), and San Vito dei Normanni (40.6N, 18.0E), Italy. For each curve, the time of the maximum *SOF* value is highlighted with a vertical line to show the temporal shift of *SOF* maximum at different locations. The relative amplitude of the maximums on the different curves is also clearly shown in the figure.

The same procedure was used to obtain similar *SOF* curves at different locations within Italy. These values were then used to draw *SOF* maps during the eclipse in the region of latitude from 36.0N to 47.5N and longitude from 6.0E to 19.0E, by a linear interpolation procedure. The *SOF* maps obtained from 8:15 UT to 11:00 UT every 15 minutes are reported in Fig. 2(a)-(l).

Furthermore, the Sun's disk was assumed to have homogeneous luminosity, ignoring any variation due to sunspots. Under these hypotheses the reduction in solar radiation due to the eclipse is assumed to be proportional to the *SOF*:

$$I_{\infty[eclipse]} = (1 - SOF) \cdot I_{\infty} \qquad (1)$$

where $I_{\infty[eclipse]}$ and $I_\infty$ are the intensity of solar radiation at the upper limit of the atmosphere in the presence or absence of the eclipse, respectively.

An estimation of the decrease in solar radiation due to the progress of the eclipse is useful in order to study the decrease in ionization in the different ionospheric layers, as will be shown in Sections 3, 5, and 6.

## 3. Effects of the eclipse on the E-layer

It is well known that the E layer behaves like a Chapman layer, and so its critical frequency can be modeled with a good approximation as:

$$f_oE_{[mod]}=[q(\chi=0, h=h_m)\cdot \cos(\chi)/\alpha]^{1/4}, \qquad (2)$$

where $q(\chi=0, h=h_m)$ is the maximum production rate, which occurs at the height $h_m$ and when the Sun is at its zenith (i.e. when the solar zenith angle is $\chi=0$), and $\alpha$ is the recombination coefficient. It should also be noted that in (2) it can be assumed:

$$q(\chi=0, h=h_m) \propto I_\infty. \qquad (3)$$

Consequently, the critical frequency $f_oE_{[eclipse]}$ of a Chapman layer, when the intensity of the ionizing radiation is decreased by a factor $(1-SOF)$, is:

$$f_oE_{[eclipse]}=f_oE_{[mod]}\cdot(1-SOF)^{1/4}. \qquad (4)$$

For relationship (2), however, the functional dependence was specified by the following empirical formula (Davies, 1965):

$$f_oE_{[mod]}=0.9\cdot[(180+1.44\cdot R_{12})\cdot \cos(\chi)]^{1/4}, \qquad (5)$$

where $R_{12}$ is the twelve-months smoothed sunspot number. As the original sunspot number has no longer been disseminated since December 2014, in this work an equivalent $R_{12}$ value was used, obtained from the empirical formula (Leitinger at al., 2005):

$$R_{12}=[167273+(F10.7-63.7)\cdot 1123.6]^{0.5}-408.99, \qquad (6)$$

where F10.7 is the solar radio flux at 10.7 cm. Relationship (6) is derived according to the ITU Recommendation (1999), in which twelve-months smoothed F10.7 values were used. In this work the daily F10.7 value for March 20, 2015 was applied, since the interest was for equivalent $R_{12}$ values suitable for the specific day of the eclipse. Besides, a comparison between modeled and observed $f_oF_1$ values showed that modeled ones obtained using equivalent $R_{12}$ values from daily F10.7 data are more consistent with observations than those obtained using equivalent $R_{12}$ values from twelve-months smoothed F10.7 data.

Fig. 3(a)-(c) shows the trends as a function of time of $f_oE_{[mod]}$, as provided by (5), and $f_oE_{[eclipse]}$ provided by (4), along with the few available experimental values scaled manually from the ionograms recorded every 15 minutes at the ionospheric stations of Rome, Gibilmanna, and San Vito dei Normanni. In many cases the cusp of the E region could not be clearly identified in the available ionograms, as a result of the combined effects of the limited efficiency of the receiving antennas at low frequencies, absorption in the D region, and the $E_s$ layer blanketing effect. To relate the trends of $f_oE_{[obs]}$, $f_oE_{[mod]}$, and $f_oE_{[eclipse]}$ to the evolution in time of the $SOF$ at each station, Fig. 3(a)-(c) also reports the relative $SOF$ curves, with the time of the maximum $SOF$ value highlighted with a vertical line. Computing the values of $\chi$ at any location for fixed times, $f_oE_{[eclipse]}$ maps were obtained during the eclipse in the region of latitude from 36.0N to 47.5N and longitude from 6.0E to 19.0E. The $f_oE_{[eclipse]}$ maps obtained every 15 minutes from 8:15 UT to 11:00 UT are reported in Fig. 4(a)-(l).

## 4. Effects of the eclipse on the sporadic E layer

A study of the $E_s$ layer during the solar eclipse was also performed, analyzing the backscatter echo received from the ionosondes of Rome and Gibilmanna. The ionosonde height-time-intensity (HTI) analysis, which was introduced to investigate $E_s$ vertical motion and variability by Haldoupis et. al (2006), was applied to study $E_s$ occurrence and behavior during the eclipse. Figs. 5(a) and 6(a) report the analysis of echoes received in the 3 days before and the 3 days following the eclipse, as well as on the day of the eclipse, while Figs. 5(b) and 6(b) report the 10 days before and 10 days after the solar eclipse. The recorded values are shown in a light blue color scale, proportional to the energy received by the ionosondes in the 3.8 – 4.2 MHz band, averaged over the time periods considered. The brightest light blues represent the highest mean energy values. The purple points in the figures represent the values recorded on March 20, 2015, the day of the solar eclipse. The vertical lines on the plots represent respectively the start and end times of the solar eclipse itself.

## 5. Effects of the eclipse on the F1-layer

The semi-empirical DuCharme et al. (1971, 1973) formula is often used to calculate the value of the critical frequency of the F1 layer, $f_oF_1$. The DuCharme et al. formula was calculated using a database from 1954 through 1966 with 39 selected stations. This formulation assumes limits for the presence of $f_oF_1$ as a function of $\chi$ and the solar activity given by the $R_{12}$ index. These limit assumptions imply that that $f_oF_1$ ionization is never present during winter or during nighttime. The trend of the critical frequency $f_oF_1$ is thus described by the following formula:

$$f_oF_{1[mod]} = f_s[\cos(\chi)]^\eta \qquad \text{for } \chi \leq \chi_m. \tag{7}$$

In this relationship:

$$\eta = 0.093 + 0.0046 \cdot \lambda_m - 0.000054 \cdot \lambda_m^2 + 0.0003 \cdot R_{12}, \tag{8}$$

and:

$$f_s = f_0 + [(f_{100} - f_0) \cdot (R_{12}/100)], \tag{9}$$

where $f_0 = 4.350 + 0.0058 \cdot \lambda_m - 0.000120 \cdot \lambda_m^2$, $f_{100} = 5.348 + 0.0110 \cdot \lambda_m - 0.000230 \cdot \lambda_m^2$, and $\lambda_m$ is the geomagnetic latitude. The maximum solar zenithal angle $\chi_m$ for which the $F_1$ layer is assumed present depends upon $R_{12}$ and $\lambda_m$.

Using this relationship, daytime $f_oF1$ can be calculated for a particular geographic position, taking into account the solar activity and geomagnetic coordinates of the site.

The validity of the DuCharme et al. formula for predicting values of $f_oF_1$, including zenith angles beyond the limits specified in the original paper, was confirmed by Scotto et al. (1997).

When an extraordinary event, such as a solar eclipse, occurs, this formula is expected not to be able to describe the $f_oF_1$ variations. The *SOF* was introduced in the DuCharme et al. formula as a correction factor in order to describe the decrease in $f_oF_1$ during such events. Hence, the following formula for $f_oF_1$ estimation during an eclipse was proposed:

$$f_oF_{1[eclipse]} = f_oF_{1[mod]} \cdot (1 - SOF)^\eta, \tag{10}$$

in accordance to the law expressed by the DuCharme et al. formulation (7) for $f_oF_1$. To describe the behavior of $f_oF_1$ during the solar eclipse the values manually scaled from the ionograms recorded every 15 minutes at the ionospheric stations of Rome, Gibilmanna, and San Vito dei Normanni were compared with:

a) the values of $f_oF_1$ provided by the DuCharme et al. formula which, instead of $R_{12}$, used the equivalent $R_{12}$ values obtained from (6) using F10.7 for March 20, 2015;

b) the values of $f_oF_1$ provided by the DuCharme et al. formula specified in point a), corrected according to (10).

The values for this comparison are reported in Fig. 7(a)-(c), along with the evolution in time of the *SOF* at each station, where the time of the maximum *SOF* value is highlighted with a vertical line. Fig. 8(a)-(l) shows the $f_oF_{1[eclipse]}$ maps obtained for the sounding hours in the region of latitude from 36.0N to 47.5N and longitude from 6.0E to 19.0E, with the computation of $\chi$ in the area considered for the fixed times.

**6. Effects of the eclipse on the F2-layer**

Fig. 9(a)-(c) uses red dots to show the time behavior of the observed values of the critical frequency of the F2 layer, $f_oF_{2[obs]}$ at the ionospheric stations of Rome, Gibilmanna, and San Vito dei Normanni. $f_oF_{2[obs]}$ are the measured $f_oF_2$ values deduced from the ionograms by manual scaling. The same figure uses black dots to show $f_oF_{2[mod]}$, which are the values representing the temporal evolution of $f_oF_2$, as would be expected in the absence of the eclipse. The $f_oF_{2[mod]}$ values are derived by reducing the monthly median by a constant factor. This factor is calculated so as to minimize the root-mean-square deviations (*RMSD*s) between the observed values and those modeled, considering the ionograms recorded before and after the eclipse.

Next, the differences $d=f_oF_{2[mod]}-f_oF_{2[obs]}$ at the ionospheric stations of Rome and Gibilmanna were calculated and are reported with black dots in Fig. 10 as a function of the *SOF*. A quadratic regression was applied to obtain $d=d(SOF)$. The relationship obtained is:

$d(SOF)=5.5 \cdot SOF^2 - 5.4 \cdot SOF$. (11)

In Fig. 10, the relationship (11) was also plotted, along with the differences $d=f_oF_{2[mod]}-f_oF_{2[obs]}$ calculated at the ionospheric station of San Vito dei Normanni and reported with red dots. These values were used to test the relationship (11) itself. An *RMSD* of 0.21 MHz was obtained, which is indicative of a close agreement with these data. In order to take into account the effect of the eclipse on $f_oF_2$ estimation, the following formula was thus proposed:

$f_oF_{2[eclipse]}=f_oF_{2[mod]}-d(SOF)$. (12)

Fig. 11(a)-(c) shows the same data reported in Fig. 10, in terms of $d$ versus *SOF* values obtained at the three stations separately. For each reported datum is also shown the time for which the computation has been performed.

Fig. 12(a)-(c) shows the trends as a function of time of $f_oF_{2[obs]}$, $f_oF_{2[mod]}$, and $f_oF_{2[eclipse]}$ at the ionospheric stations of Rome, Gibilmanna, and San Vito dei Normanni, along with the evolution in time of the *SOF* at each station, with the time of the maximum *SOF* value highlighted with a vertical line.

Fig. 13(a)-(l) shows the $f_oF_{2[eclipse]}$ maps obtained every 15 minutes in the region of latitude from 36.0N to 47.5N and longitude from 6.0E to 19.0E. These maps were obtained applying the relationship (12) to the entire geographical area, using the *SOF* maps calculated in Section 2, and the monthly median $f_oF_2$ values mapping procedure performed by the IRI model (Jones and Gallet, 1962; Jones et al., 1969).

**7. Behavior of a regional 3D ionospheric model during the eclipse.**

In this study the behavior of a regional adaptive and assimilative 3D ionospheric model (Sabbagh et al., 2016) was also tested while assimilating plasma frequency profiles $f_p(h)$ during the solar eclipse. The 3D model ingests ionosonde data by minimizing the *RMSD*s between the observed and modeled values of $f_p(h)$ profiles obtained at the points where the observations are available. This minimization procedure is used to set the values of some free parameters that characterize the model. These values are then applied over the specified area.

In this study the model was applied over the region of latitude from 36.0N to 47.5N and longitude from 6.0E to 19.0E. The model performance was evaluated in terms of adaptability to ionospheric conditions observed at a given moment over Rome and Gibilmanna, while the accuracy with which the algorithm is able to model $f_p$ was tested over San Vito dei Normanni.

The model input $f_p(h)$ profiles at Rome and Gibilmanna were obtained from Autoscala (Scotto and Pezzopane, 2002; Pezzopane and Scotto, 2005) using the *Adaptive Ionospheric Profiler* (AIP) (Scotto, 2009) as part of the vertical ionogram automatic interpretation process. The Rome and Gibilmanna ionograms were recorded by the AIS-INGV ionosonde (Zuccheretti et al., 2003). The formulations (4) and (10) were introduced into Autoscala in order to model $f_oE$ values and set the $f_oF_1$ search procedure in the ionograms, while $f_oF_2$ values are autoscaled using an image recognition technique.

The San Vito dei Normanni ionograms were recorded by the DPS-4 Digisonde (Hanes, 1994) and inverted by POLAN (Titheridge, 1959, 1985, 1988), using manually scaled $f_oF_1$ and $f_oF_2$ values. $f_oE$ POLAN input values were modeled by (4), as it was not possible to clearly identify the cusp of the E region in the ionograms.

The *SOF* corrections (4), (10), and (12) for the critical frequency evaluations were introduced into the model so that it was able to follow the eclipse effects. In the previous sections some model outputs during the eclipse are shown. The $f_oE$ and $f_oF_1$ values used by the model as $f_p(h)$ anchor point coordinates are not affected by the adaptation procedure, and so Figs. 4(a)-(l) and 8(a)-(l) really represent critical frequency maps for the E and F1 layers generated by the model itself.

In contrast, the $f_oF_2$ values computed by (12) were used by the model as empirical values to be adapted to actual conditions. Through the adaptation procedure mentioned above, the value was set for the $f_oF_2$ variation, $\Delta f_oF_2$, to be applied over the entire area as an added factor. Then, the $f_oF_2$ model output values are computed by:

$$f_oF_2 = f_oF_{2[eclipse]} + \Delta f_oF_2, \tag{13}$$

where $f_oF_{2[eclipse]}$ are computed by (12), and $\Delta f_oF_2$ is the constant value obtained from the adaptation procedure performed at the Rome and Gibilmanna ionospheric stations.

Fig. 14(a)-(l) shows the $f_oF_2$ maps obtained every 15 minutes for the specified region using the procedure described above. It is noteworthy that the shapes of the $f_oF_2$ isolines in Fig. 14(a)-(l) are the same as the corresponding isolines in Fig. 13(a)-(l), since the $f_oF_2$ values in the two cases are only shifted by a constant factor.

When the model is unable to adapt to actual conditions, it can provide a description of a "quiet" ionosphere based on empirical monthly median values for its descriptive parameters, such as critical frequencies. For cases of non adaptation during the eclipse, Fig. 14 reports maps of $f_oF_{2[eclipse]}$ values computed by (12) (i.e. maps of $f_oF_2$ values computed by (13) with $\Delta f_oF_2=0$). In these cases the $f_oF_{2[eclipse]}$ maps in Fig. 14 are the exact same as those reported in Fig. 13. These specific cases are (c), (h), (i), and (k), corresponding to 08:45 UT, 10:00 UT, 10:15 UT, and 10:45 UT hours, respectively.

As an example, Figs. 15(a)-(c), 16(a)-(c), and 17(a)-(c) report other model outputs obtained in the specified region for March 20, 2015 at 09:30 UT. In particular, Fig. 15(a)-(c) shows $f_p$ maps at the fixed heights of 110 km, 180 km, and 300 km, while Fig. 16(a)-(c) provides cross-sectional maps at the fixed latitudes of 37.5N, 41.75N, and 46N, with $h$ from 60 km to $h_mF_2$, while Fig. 17(a)-(c) provides cross-sectional maps at the fixed longitudes 8E, 12.5E, 17E with $h$ from 60 km to $h_mF_2$.

## 8. Conclusions

This study investigated the effect of the solar eclipse of March 20, 2015 on the ionosphere. The eclipse was characterized by the SOF time evolution at any point over the Earth's surface, in order to take into account the decreased solar irradiance during the eclipse.

The *SOF* values were obtained from a simplified model, and used to propose corrective formulations for estimation of the E, F1, and F2 layer critical frequencies during the eclipse. In these formulations equivalent $R_{12}$ values obtained from daily F10.7 values were used, since the interest was for the specific day of the eclipse.

As shown in Figs. 3(a)-(c), 7(a)-(c), and 12(a)-(c), critical frequency values computed by the corrective formulations (4), (10), and (12) provide close agreement with experimental values obtained from vertical ionospheric soundings performed at the Rome, Gibilmanna, and San Vito dei Normanni stations, since they follow the *SOF* curve trends in counter-phase at the relative locations. The capability of these formulations to follow the eclipse effects is also shown in Figs. 4(a)-(l), 8(a)-(l), and 13(a)-(l), which report critical frequency maps for the region of latitude from 36.0N to 47.5N and longitude from 6.0E to 19.0E obtained every 15 minutes. As can be seen, the isolines shown in these figures match the shape of the *SOF* isolines shown in Fig. 2.

Concerning the $E_s$ layer behavior, when ionogram analysis is limited to 3 days before and 3 days after the eclipse, the appearance of the $E_s$ layer seems to be related to the event (Figs. 5(a) and 6(a)), since its presence is not evident during the period of a few days centered on the day of the eclipse. However, when a more extensive set of days before and after the eclipse are taken into account this phenomenon does not appear so clear (Figs. 5(b) and 6(b)), because echoes from the same virtual heights of reflection are clearly present, averaging over this longer time period. Furthermore, the HTI analysis highlights the descent in altitude of the $E_s$ layer with the characteristic tidal semidiurnal periodicity. These results are not in line with what is reported by most authors, who suggest an obvious relationship between $E_s$ layer occurrence and solar eclipses, as mentioned in Section 1.

Fig. 12(a)-(c) shows that the minimum for $f_oF_{2[obs]}$ approximately coincides with the maximum *SOF* at the ionospheric stations of Gibilmanna and San Vito, and is in advance at Rome. In this study the predicted delay in the response of the F2 region to the eclipse was not observed. This may be explained as an effect of the magnetic storm that started on March 17 (St. Patrick's Day magnetic storm), which generated not insignificant dynamic forces such as thermospheric winds and electric fields. The reason for the different behavior of the ionospheric station in Rome, also obvious in Fig. 11(a)-(c), remains unclear.

The corrections (4), (10), and (12) were then introduced into a regional adaptive and assimilative 3D ionospheric model to enable it to follow the eclipse effects, improving its performances. The 3D model was applied over the region of latitude from 36.0N to 47.5N and longitude from 6.0E to 19.0E, with the ingestion of the $f_p(h)$ profiles obtained over Rome and Gibilmanna from ionograms recorded every 15 minutes during the solar eclipse.

The performance of the model in terms of adaptability to ionospheric conditions observed at any given moment were evaluated by computing the *RMSD*s between the observed and modeled values of $f_p(h)$ profiles obtained at Roma and Gibilmanna. Considering the cases in which the model demonstrated its capability to fit the observed profiles, the mean *RMSD* value obtained was 0.46 MHz when all available profiles were used, and 0.39 MHz when considering only validated profiles.

The accuracy with which the algorithm can model $f_p$ was tested over San Vito dei Normanni using the same procedure. In this case the mean *RMSD* value obtained was 0.46 MHz when all available profiles were used, and 0.43 MHz when considering only validated profiles.

All these values are in close agreement with those reported in the previous paper (Sabbagh et al., 2016), in which the degree of adaptability and accuracy of the model were assessed as quite good, compared with the best possible accuracy of critical frequency measurements from an ionosonde, equal to 0.1 MHz according to the International Union of Radio Science (URSI) standard (Piggott and Rawer, 1972), which assumes manually scaled values by experienced operators.

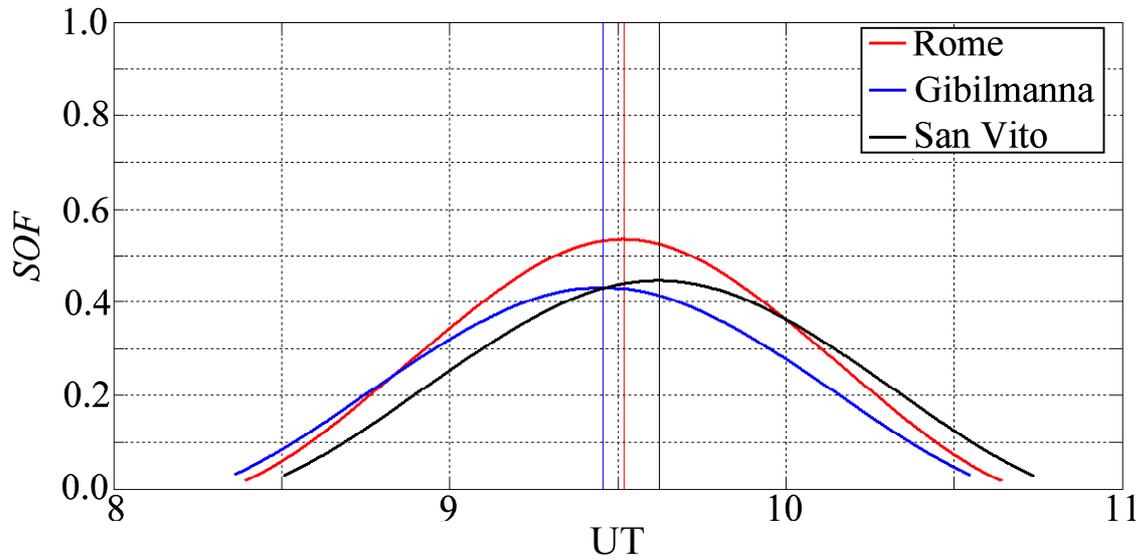

Fig. 1. Time evolution of *SOF* obtained at the ionospheric stations of Rome (41.8N, 12.5E), in red, Gibilmanna (37.9N, 14.0E), in blue, and San Vito dei Normanni (40.6N, 18.0E), in black. These curves were obtained using a simplified model which considers the motion of the Sun and Moon to be straight and uniform with both bodies circular in shape, while ignoring any variation in luminosity due to sunspots and the limb darkening effect. For each curve, the time of the maximum *SOF* value is highlighted with a vertical line.

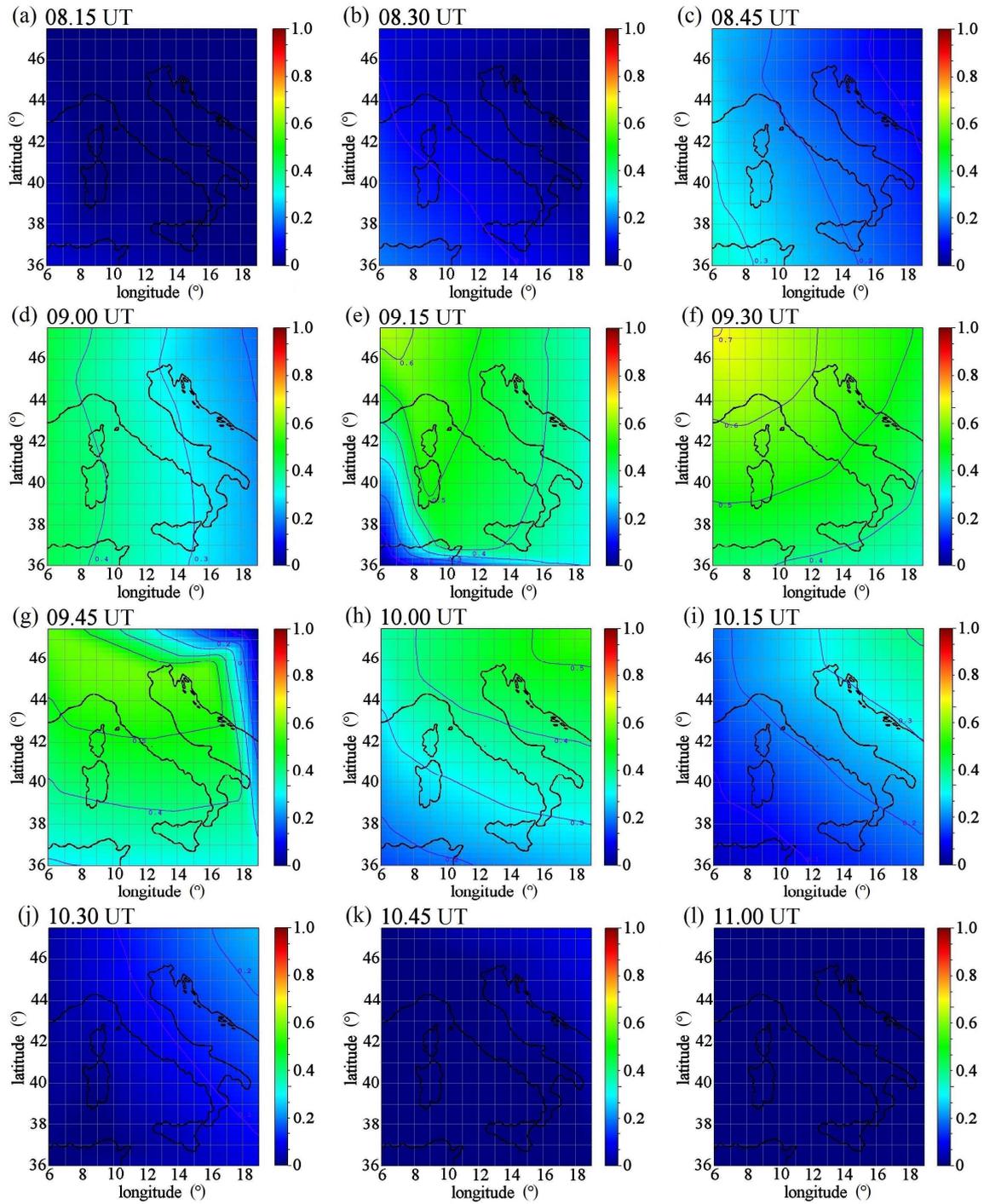

Fig. 2(a)-(l). Maps of *SOF* in the region of latitude from 36.0N to 47.5N and longitude from 6.0E to 19.0E, calculated from 8:15 UT to 11:00 UT, during the eclipse of March 20, 2015. These maps were obtained by a linear interpolation in space of the *SOF* values computed at different locations in the area considered.

● $f_oE_{[obs]}$   ○ $f_oE_{[mod]}$   — $f_oE_{[eclipse]}$   — SOF

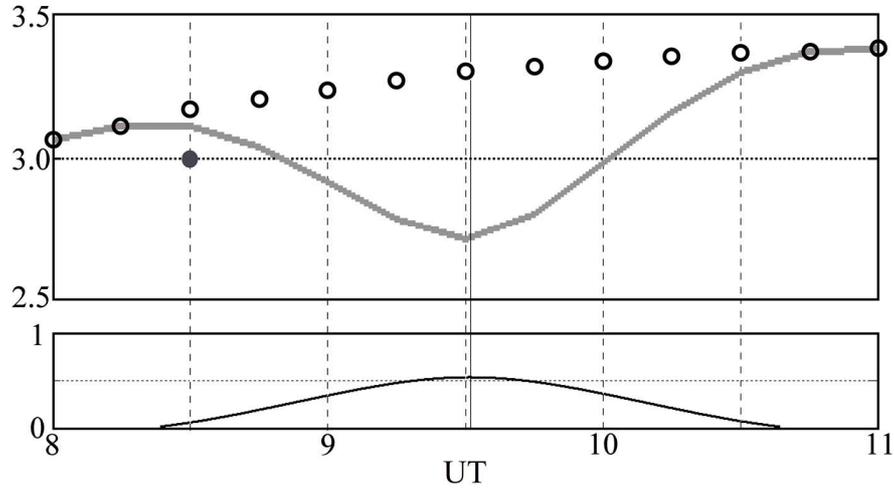

a) Rome

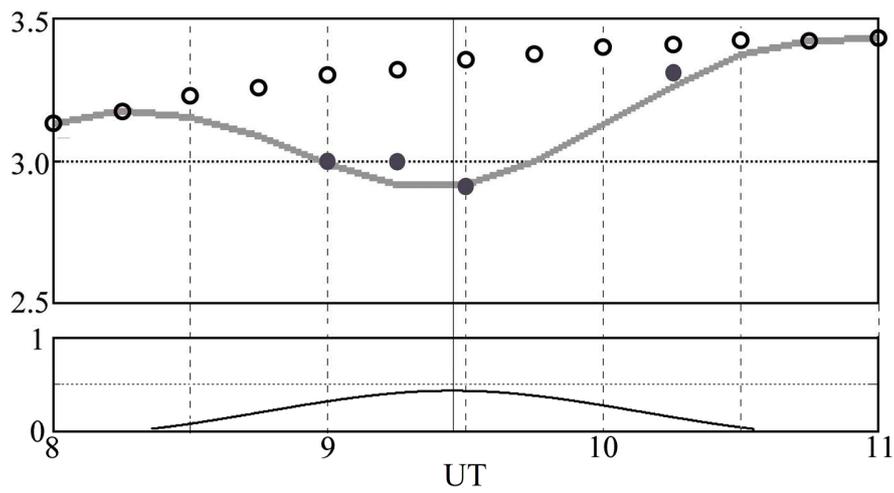

b) Gibilmanna

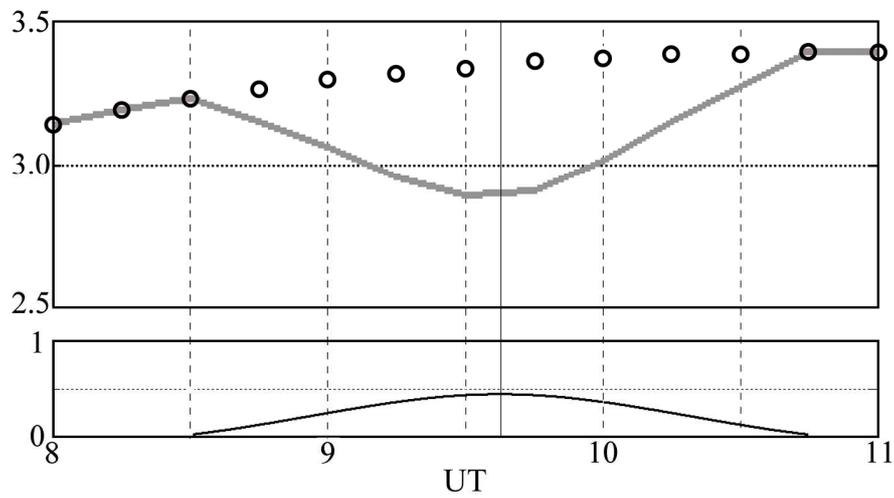

c) San Vito

Fig. 3(a)-(c). Trends as a function of time from 8:00 UT to 11:00 UT of $f_oE_{[obs]}$, $f_oE_{[mod]}$, and $f_oE_{[eclipse]}$ at the ionospheric stations of Rome (41.8N, 12.5E), Gibilmanna (37.9N, 14.0E), and San Vito dei Normanni (40.6N, 18.0E) during the eclipse of March 20, 2015. The trends are related to the evolution in time of the *SOF* at each station, and the time of the maximum *SOF* value is highlighted with a vertical line. In many cases the cusp of the E region could not be clearly identified from the available ionograms and so only a few $f_oE_{[obs]}$ values are available.

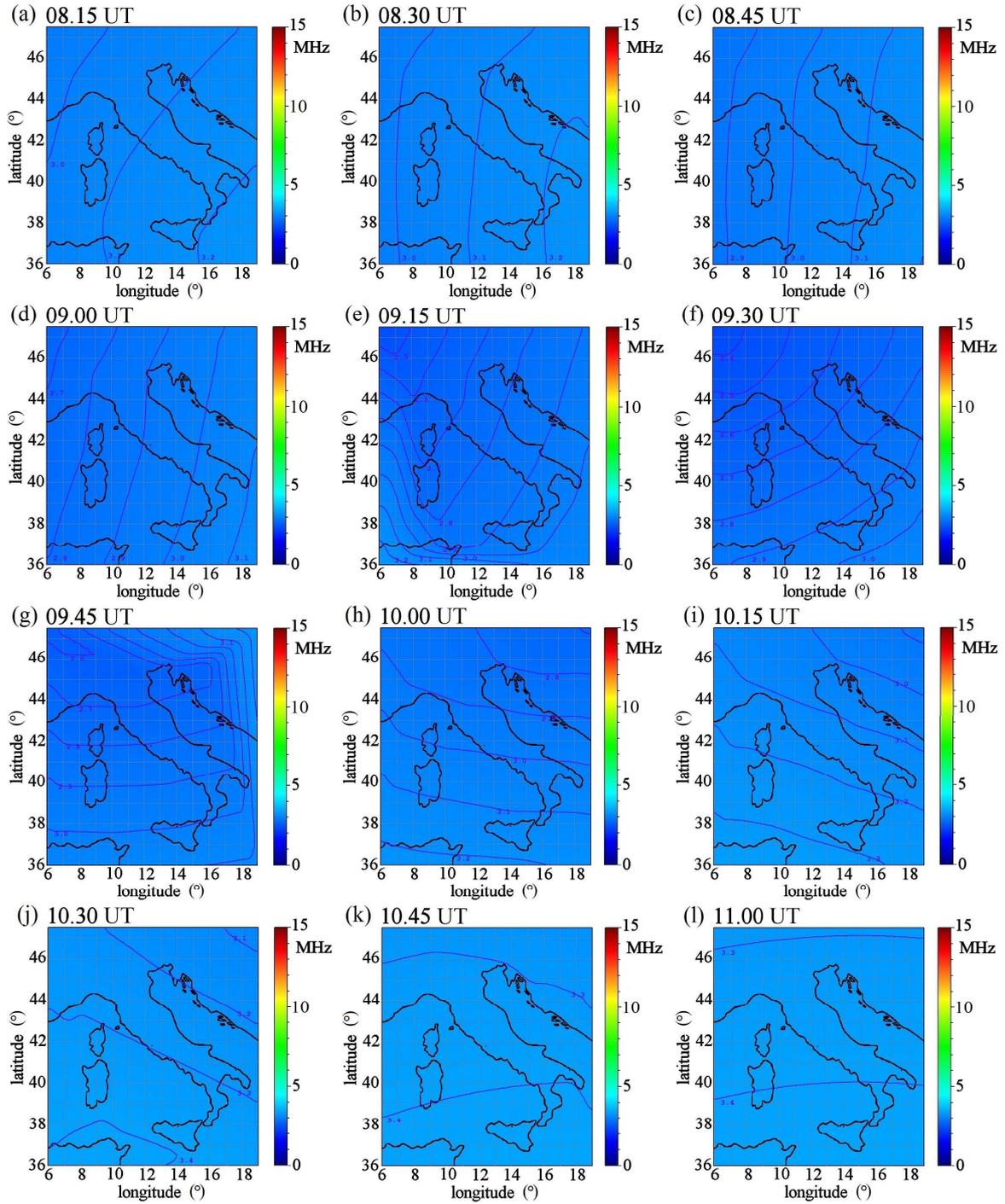

Fig. 4 (a)-(l). Maps of $f_oE_{[eclipse]}$ in the region of latitude from 36.0N to 47.5N and longitude from 6.0E to 19.0E, calculated from 8:15 UT to 11:00 UT during the eclipse of March 20, 2015. These maps were obtained by computing the values of $\chi$ every 15 minutes at any location in the area considered. .

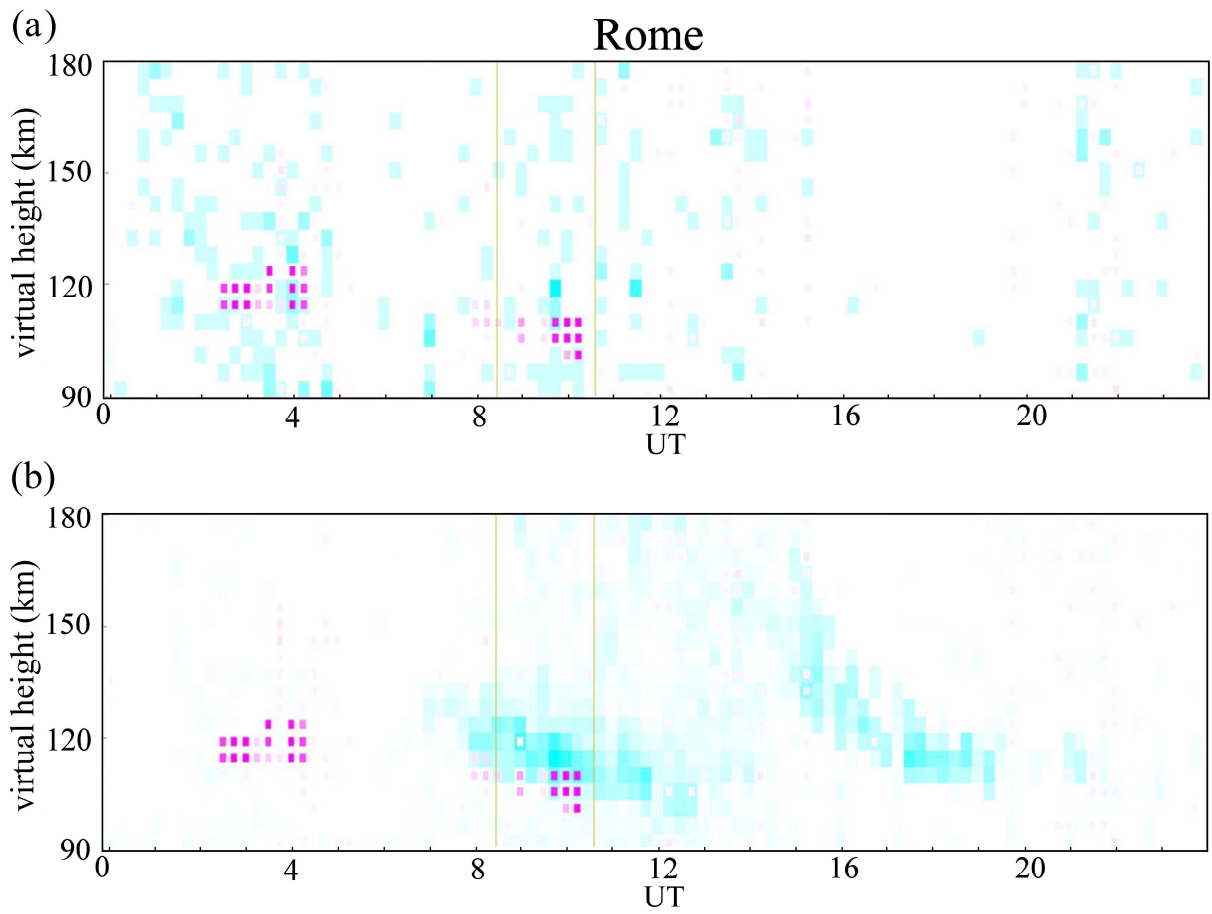

Fig. 5(a)-(b). HTI analysis applied to ionogram recordings made at the ionospheric station of Rome (41.8N, 12.5E) during the period centered on March 20, in which the eclipse occurred. The results of the 7 day (March 17 - March 23) analysis are reported in panel (a), and the corresponding 21 day (March 10 - March 31) analysis in panel (b).

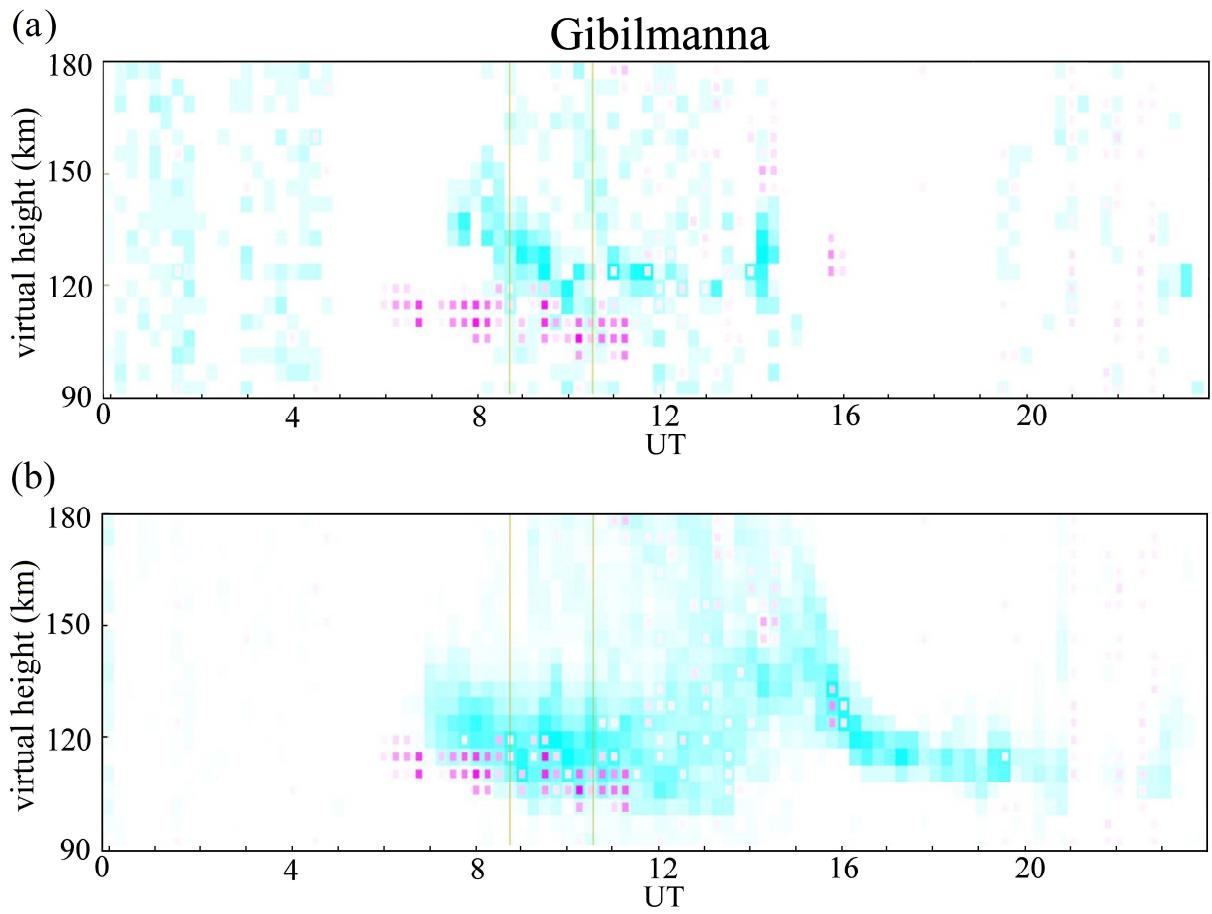

Fig. 6(a)-(b). HTI analysis applied to ionogram recordings made at the ionospheric station of Gibilmanna (37.9N, 14.0E) during the period centered on March 20, in which the eclipse occurred. The results of the 7 day (March 17 - March 23) analysis are reported in panel (a), and the corresponding 21 day (March 10 - March 31) analysis in panel (b).

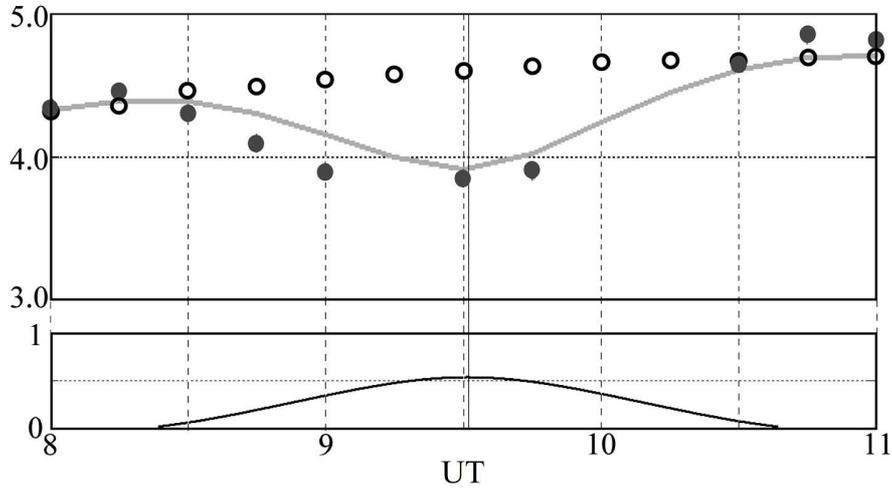
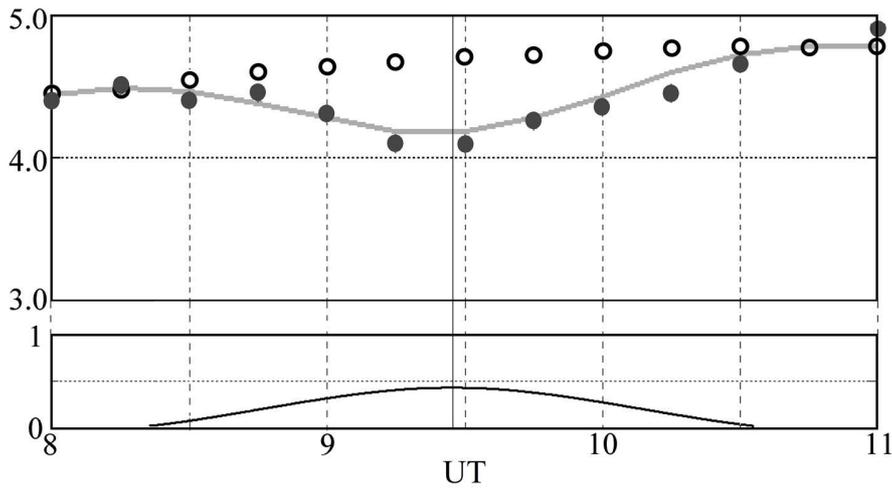
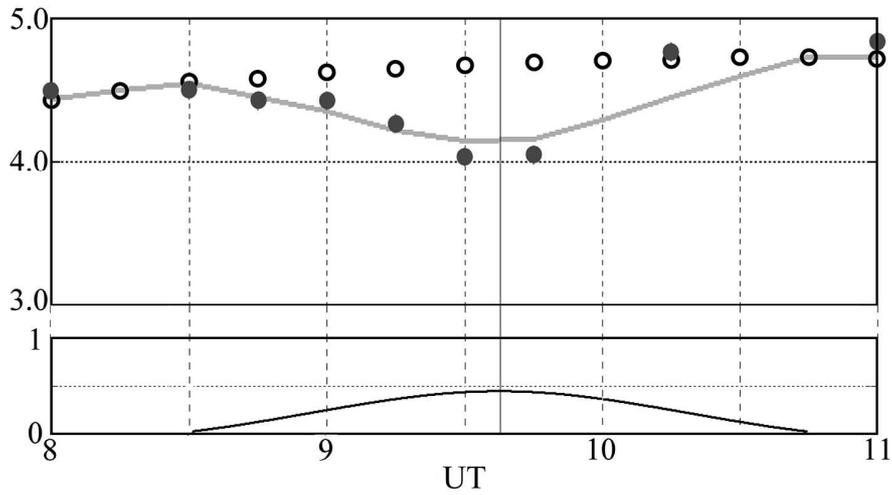

Fig. 7(a)-(c). Trends as a function of time from 8:00 UT to 11:00 UT of $f_oF_{1[obs]}$, $f_oF_{1[mod]}$, and $f_oF_{1[eclipse]}$ at the ionospheric stations of Rome (41.8N, 12.5E), Gibilmanna (37.9N, 14.0E), and San Vito dei Normanni (40.6N, 18.0E) during the eclipse of March 20, 2015. The trends are related to the evolution in time of the *SOF* at each station, and the time of the maximum *SOF* value is highlighted with a vertical line.

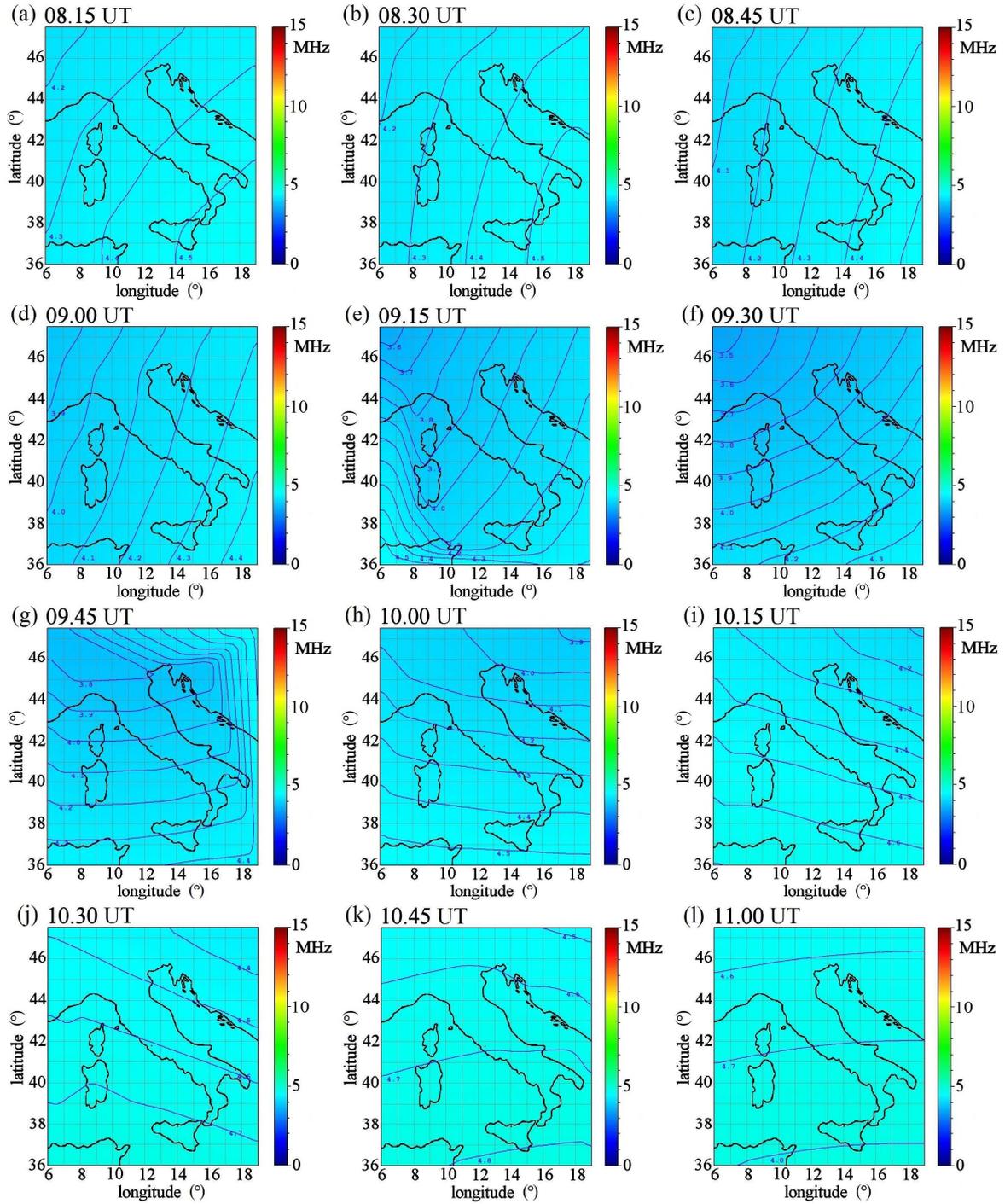

Fig. 8(a)-(l). Maps of $f_oF_{1[eclipse]}$ in the region of latitude from 36.0N to 47.5N and longitude from 6.0E to 19.0E, calculated from 8:15 UT to 11:00 UT during the eclipse of March 20, 2015. These maps were obtained by computing the values of $\chi$ every 15 minutes at any location in the area considered.

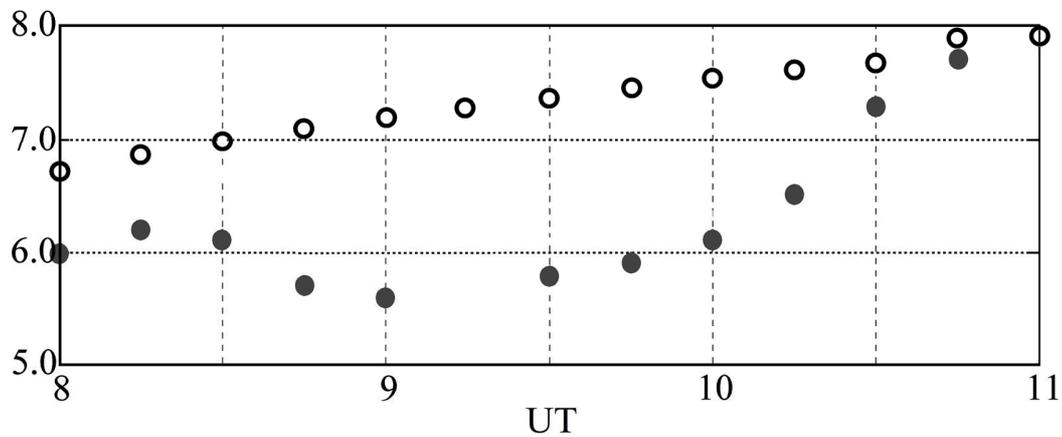

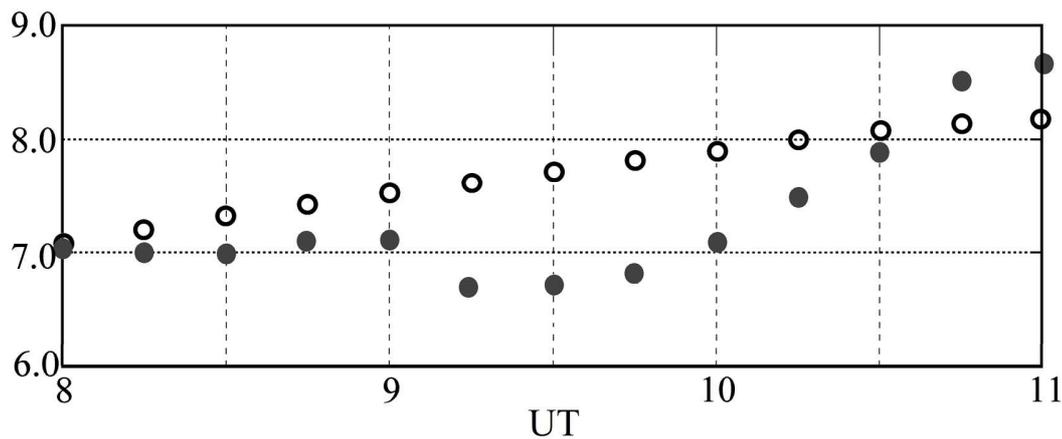

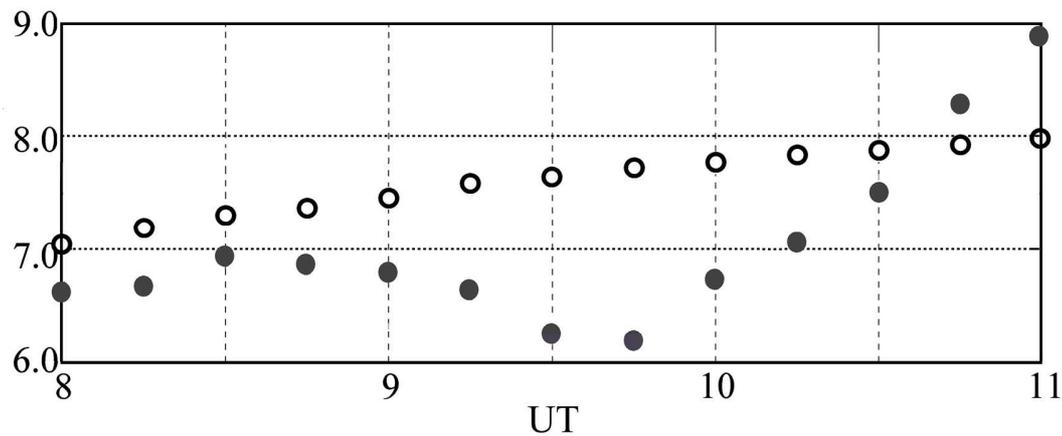

Fig. 9(a)-(c). Trends as a function of time from 8:00 UT to 11:00 UT of $f_oF_{2\,[obs]}$ and $f_oF_{2\,[mod]}$ at the ionospheric stations of Rome (41.8N, 12.5E), Gibilmanna (37.9N, 14.0E), and San Vito dei Normanni (40.6N, 18.0E) during the eclipse of March 20, 2015.

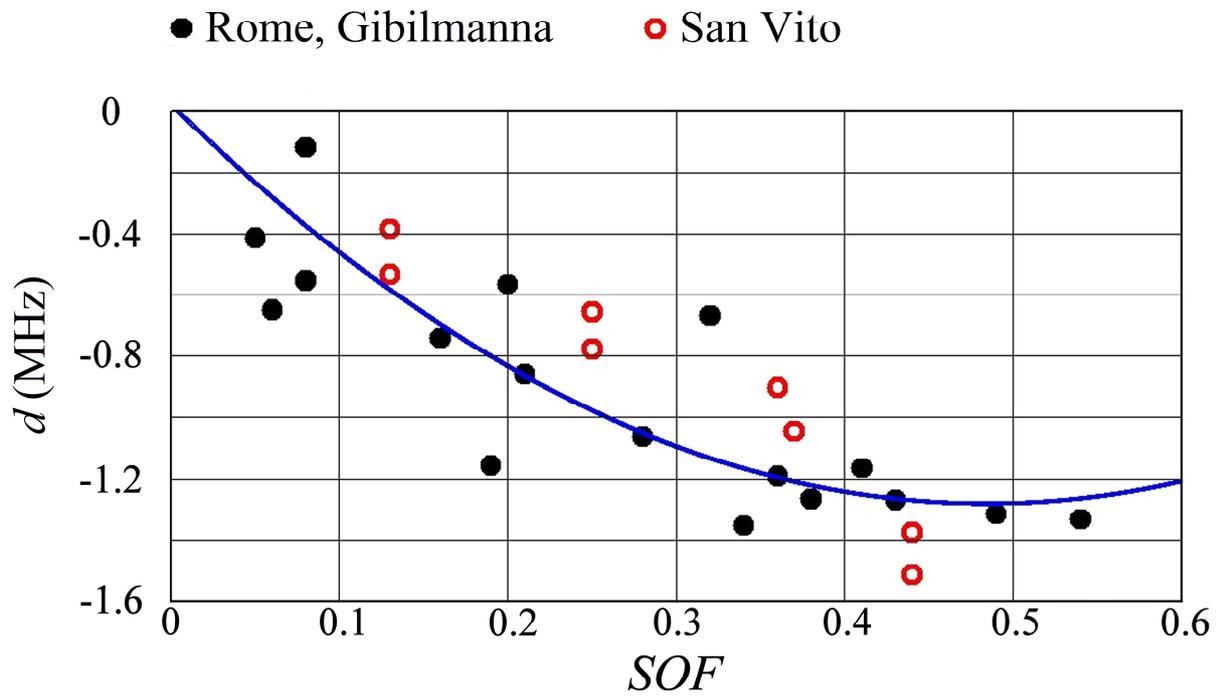

Fig. 10. Plot of *d* values computed at the ionospheric stations of Rome (41.8N, 12.5E) and Gibilmanna (37.9N, 14.0E, black dots), and San Vito dei Normanni (40.6N, 18.0E, red dots), as a function of *SOF* during the eclipse of March 20, 2015. The quadratic regression of the Rome and Gibilmanna data is also shown with a blue line. The San Vito dei Normanni data was used to test the resulting relationship.

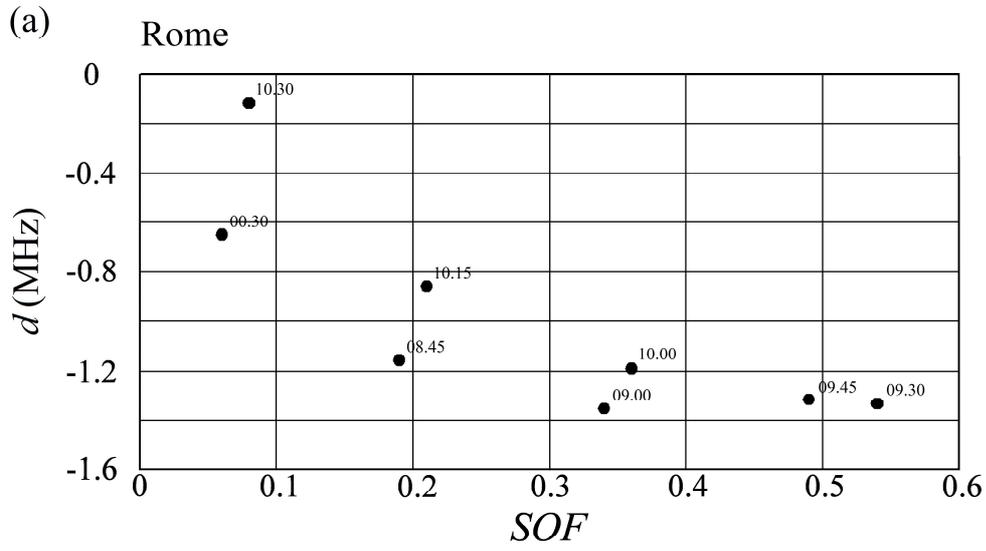
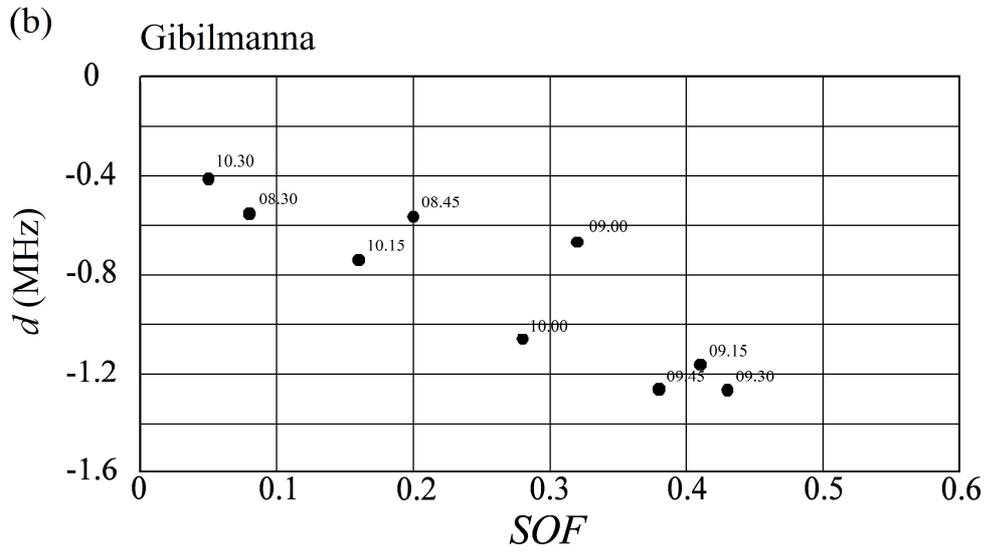
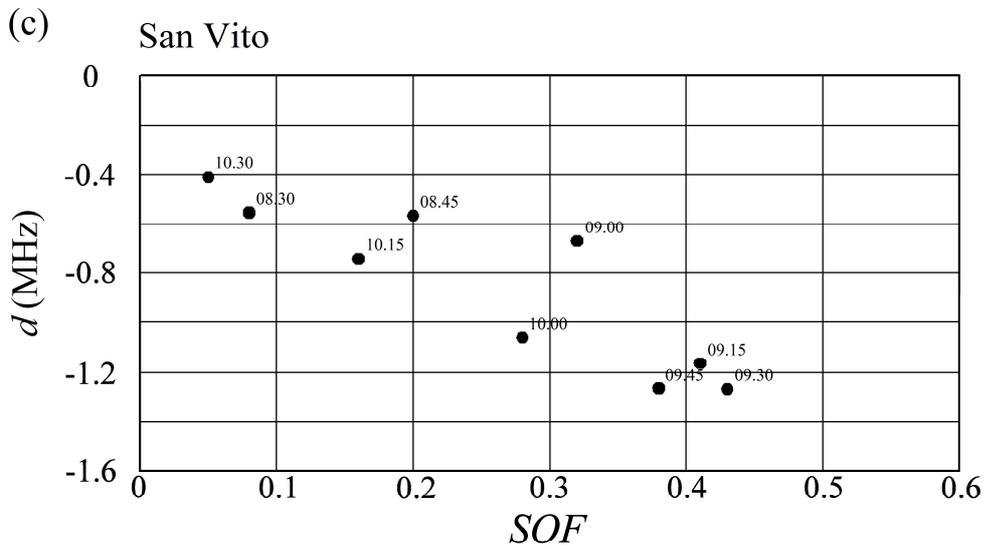

Fig. 11(a)-(c). Plot of *d* values computed at the ionospheric stations of Rome (41.8N, 12.5E), Gibilmanna (37.9N, 14.0E), and San Vito dei Normanni (40.6N, 18.0E) separately, as a function of *SOF* during the eclipse of March 20, 2015. For each reported datum is also shown the time for which the computation has been performed.

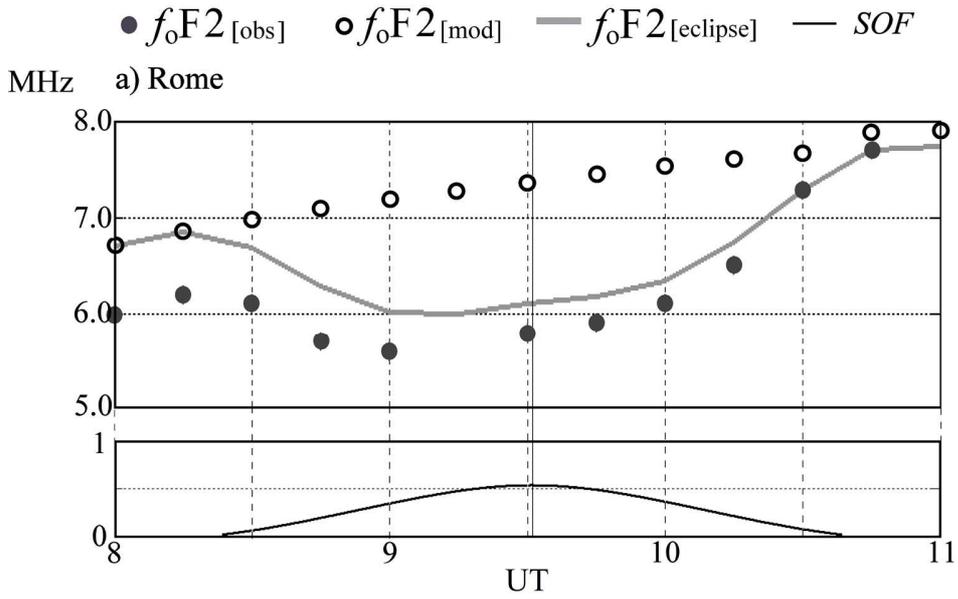
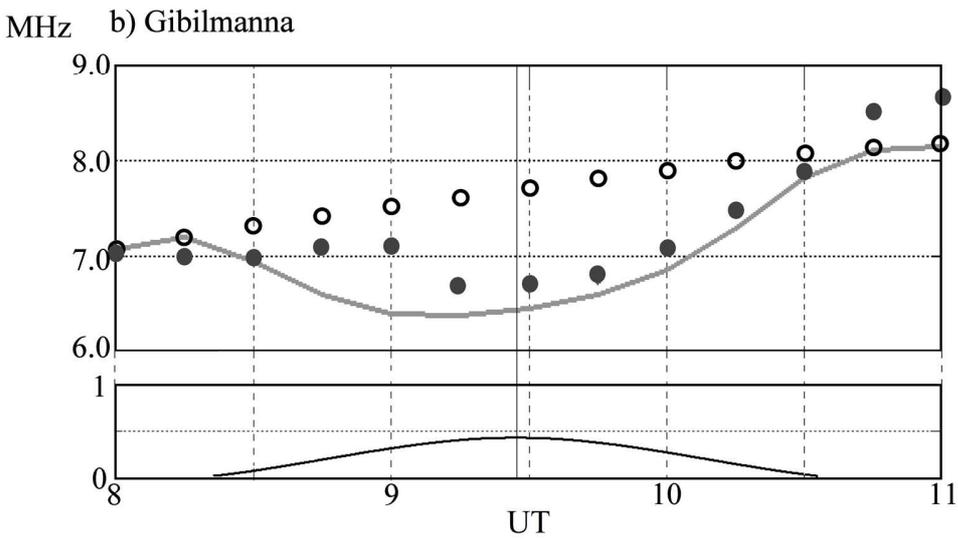
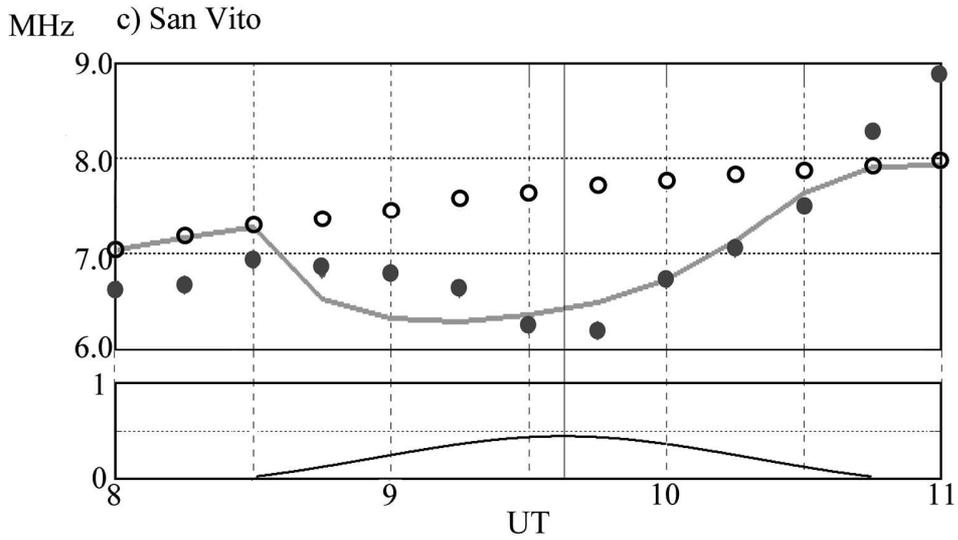

Fig. 12(a)-(c). Trends as a function of time from 8:00 UT to 11:00 UT of $f_oF_{2\,[obs]}$, $f_oF_{2\,[mod]}$, and $f_oF_{2[eclipse]}$ at the ionospheric stations of Rome (41.8N, 12.5E), Gibilmanna (37.9N, 14.0E), and San Vito dei Normanni (40.6N, 18.0E) during the eclipse of March 20, 2015. The trends are related to the evolution in time of the *SOF* at each station, and the time of the maximum *SOF* value is highlighted with a vertical line.

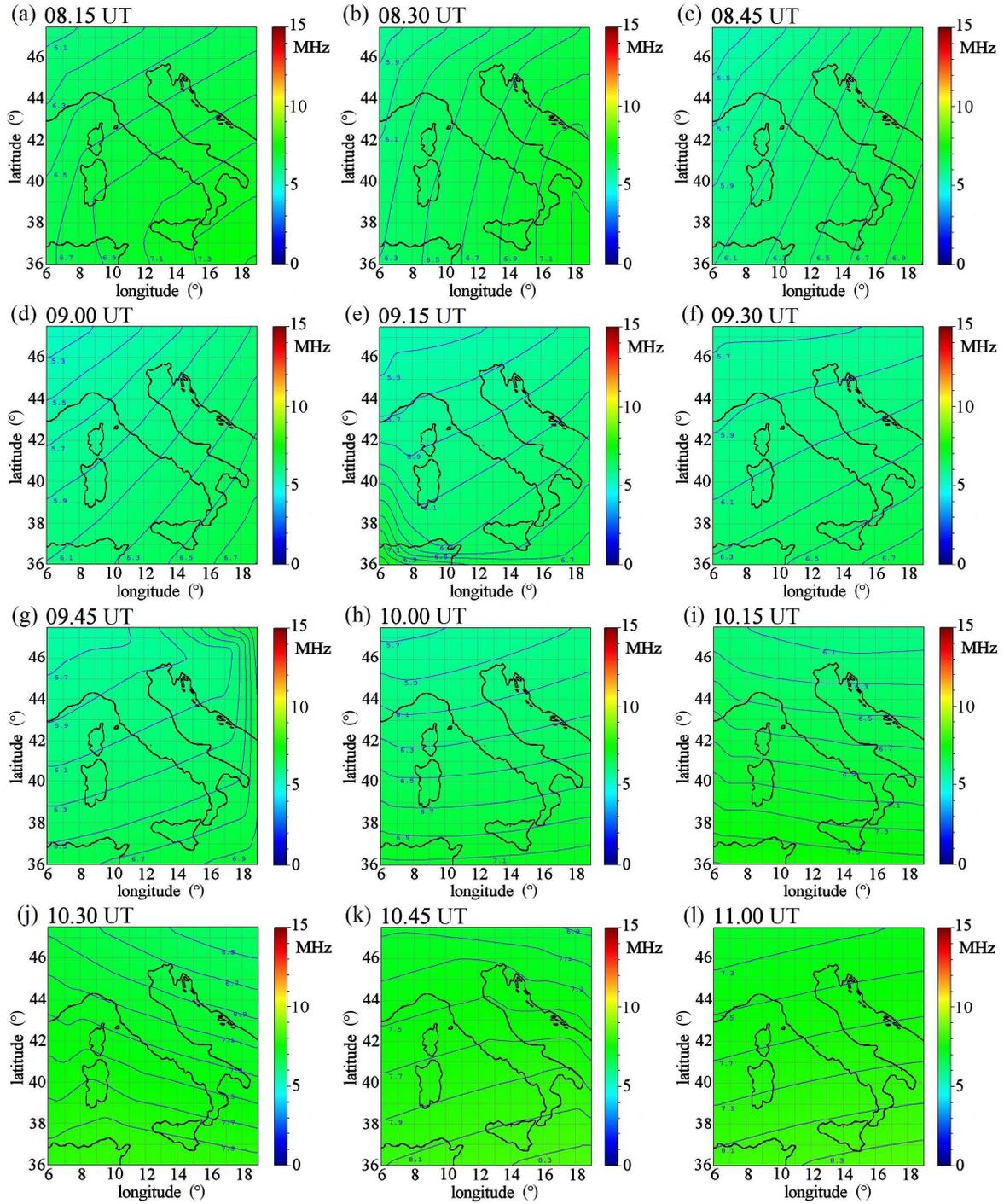

Fig. 13(a)-(l). Maps of $f_oF_{2[eclipse]}$ in the region of latitude from 36.0N to 47.5N and longitude from 6.0E to 19.0E, calculated from 8:15 UT to 11:00 UT during the eclipse of March 20, 2015. These maps were obtained using the monthly median $f_oF_2$ values mapping procedure performed by the IRI model.

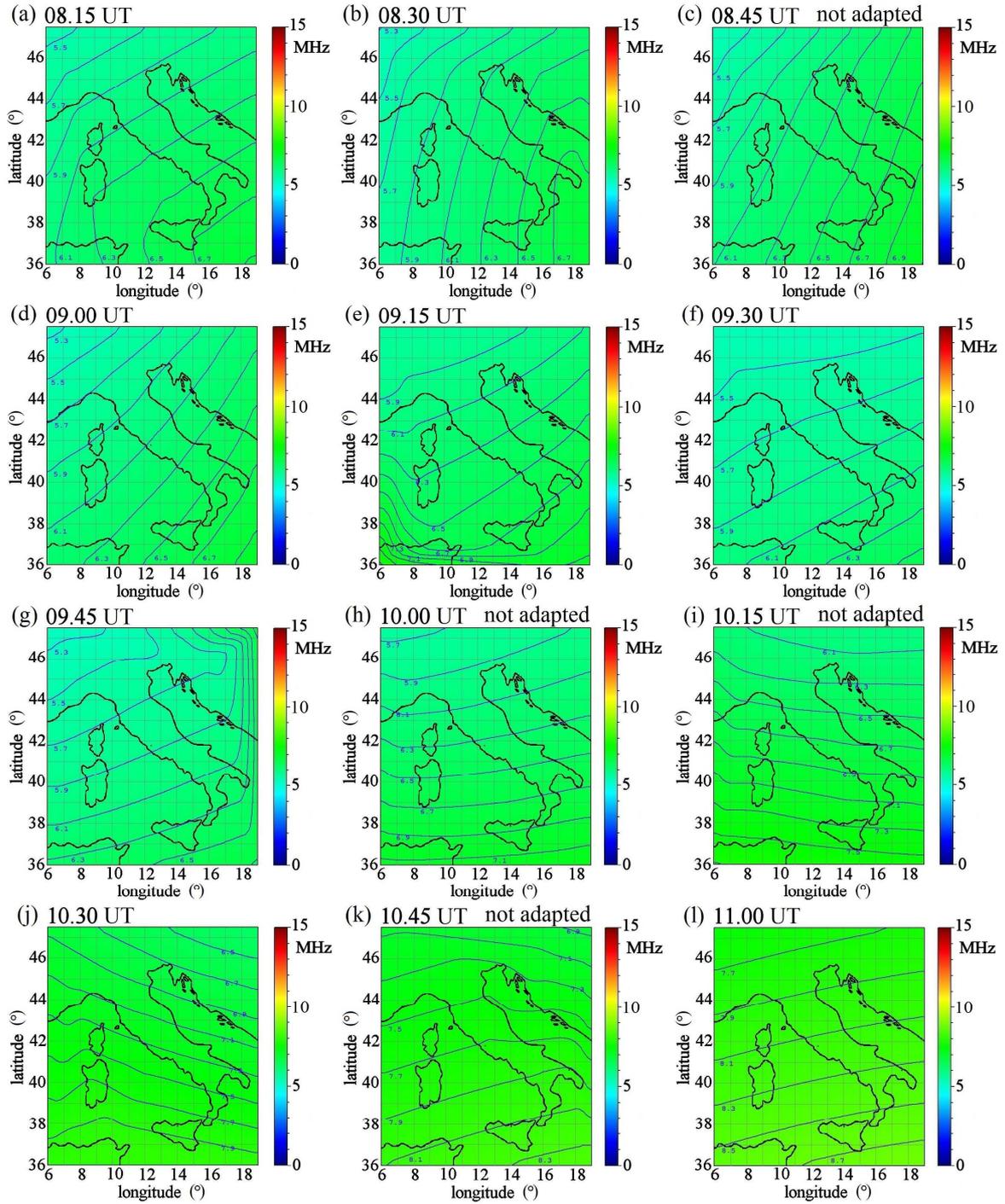

Fig. 14(a)-(l). Maps of $f_oF_2$ in the region of latitude from 36.0N to 47.5N and longitude from 6.0E to 19.0E, calculated from 8:15 UT to 11:00 UT during the eclipse of March 20, 2015. These maps were computed by a regional 3D ionospheric model which adds a constant factor $\Delta f_oF_2$ to the $f_oF_{2[eclipse]}$ values. The constant is obtained using an $f_p(h)$ adaptation procedure performed every 15 minutes at the Rome (41.8N, 12.5E) and Gibilmanna (37.9N, 14.0E) ionospheric stations. Panels (c), (h), (i), and (k) show cases in which the model was unable to adapt to actual conditions. In these cases $\Delta f_oF_2$ was set to 0.

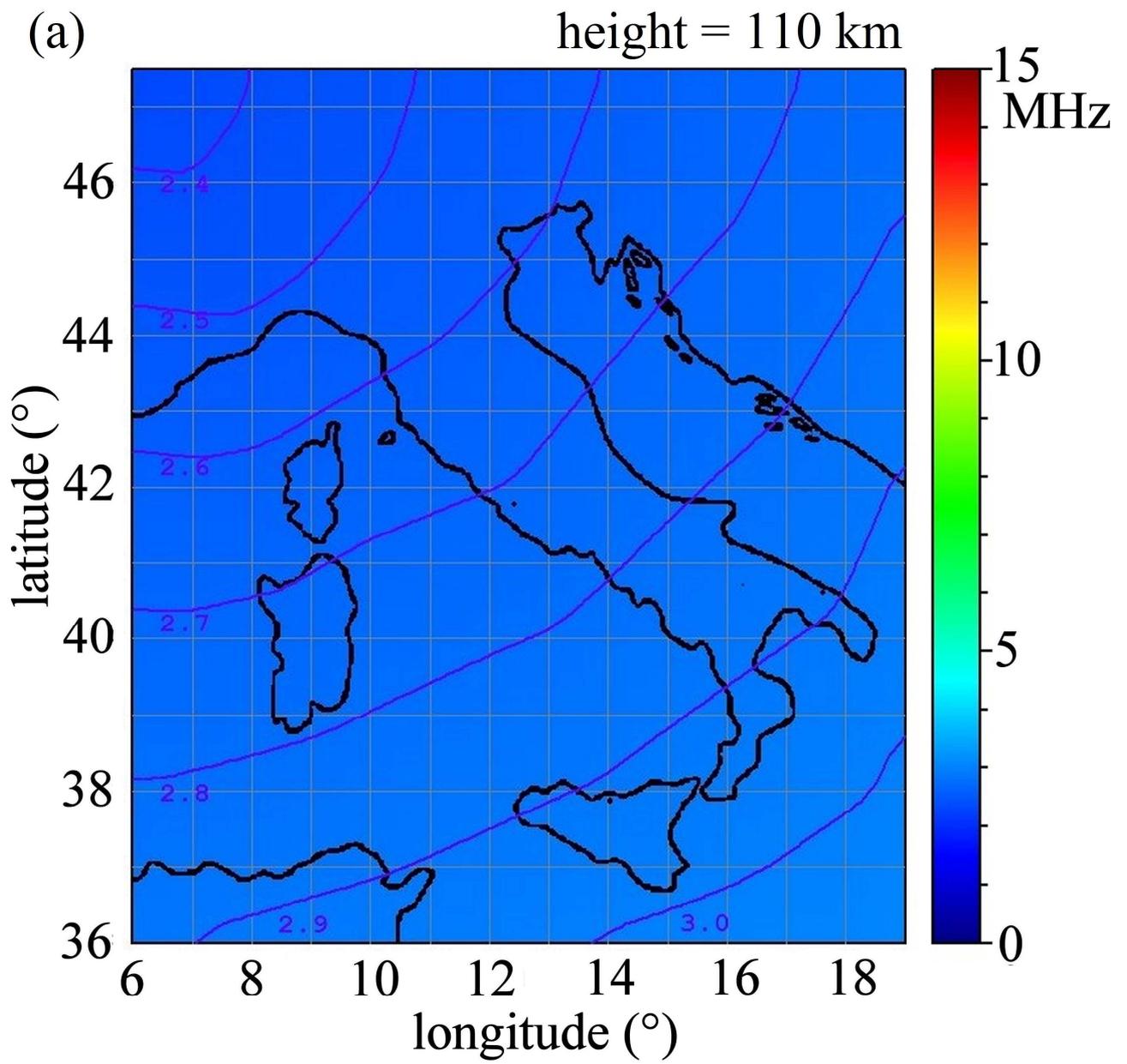

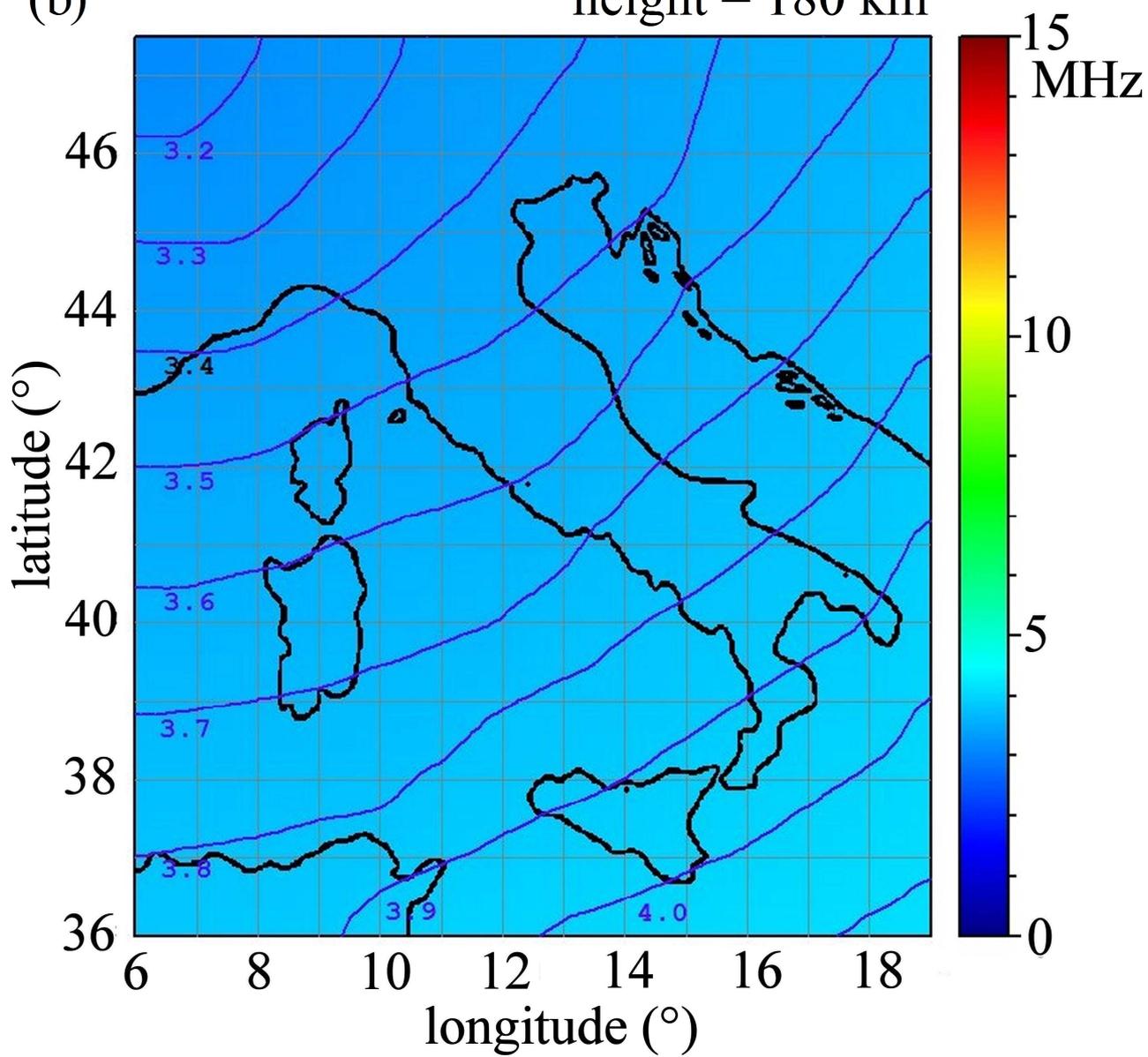

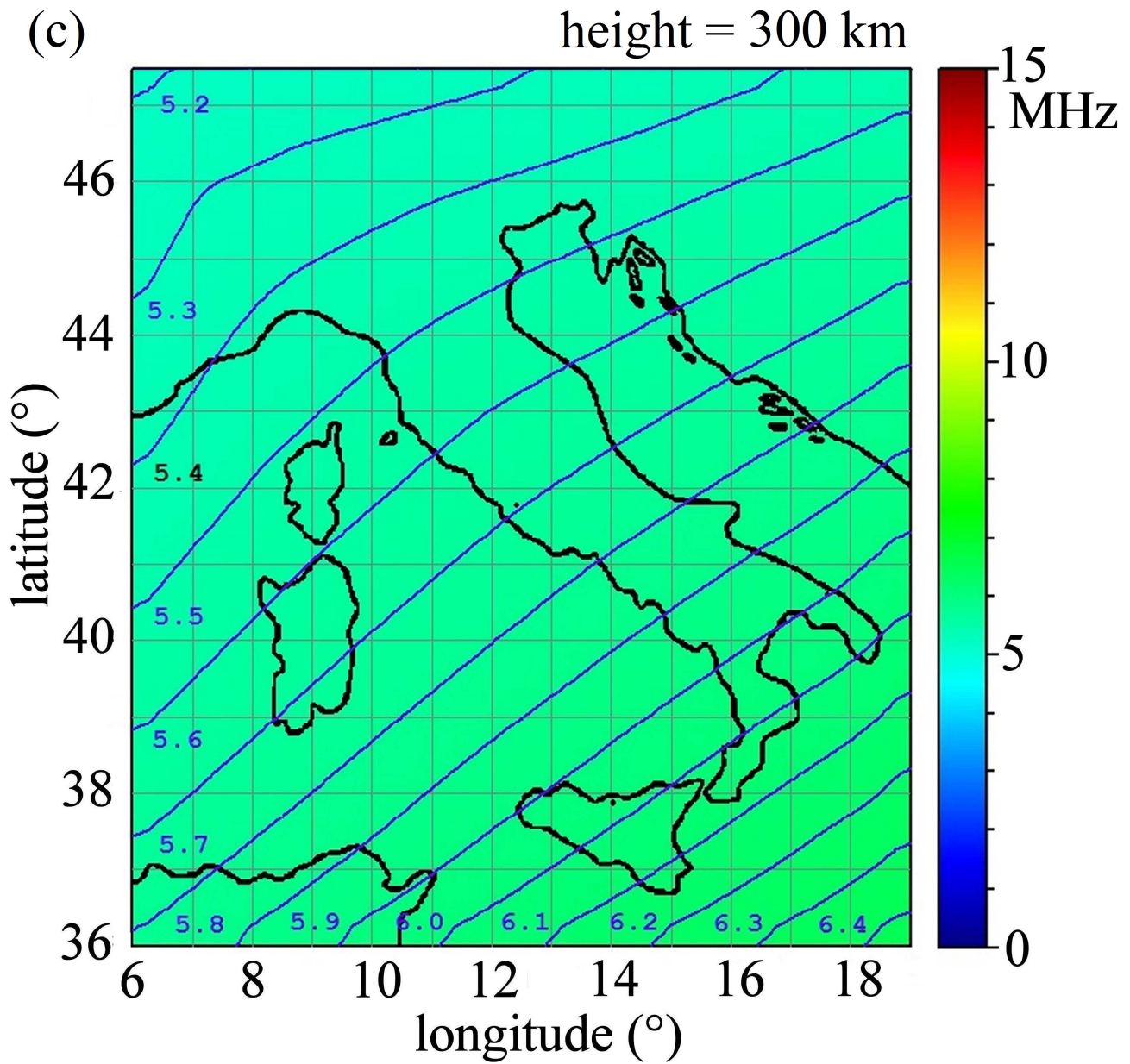

Fig. 15(a)-(c). Maps of $f_P$ at the fixed heights 110 km, 180 km, and 300 km in the region of latitude from 36.0N to 47.5N and longitude from 6.0E to 19.0E, generated for March 20, 2015 at 09:30 UT by the 3D model studied in this work.

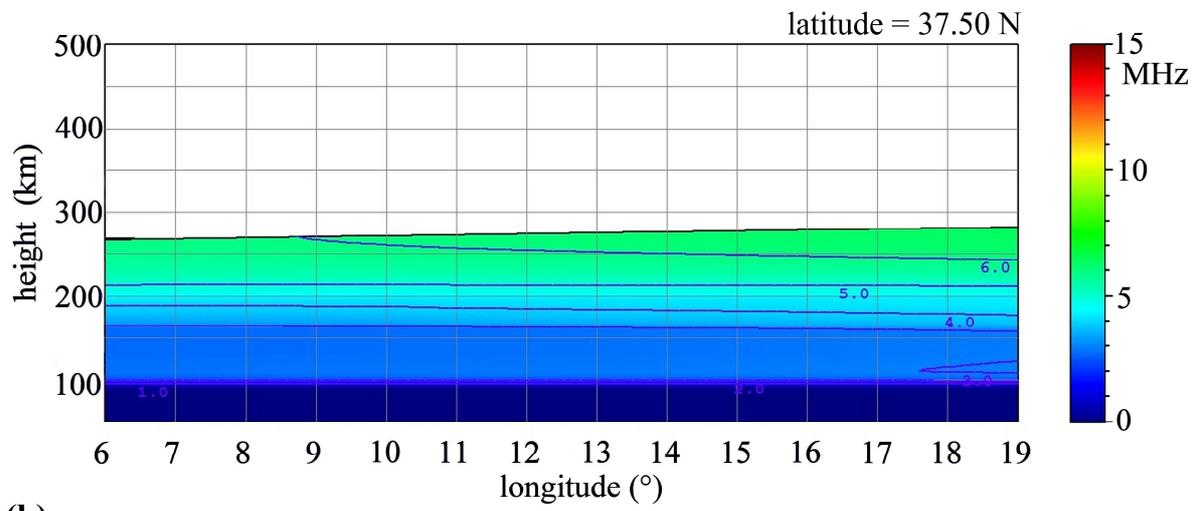

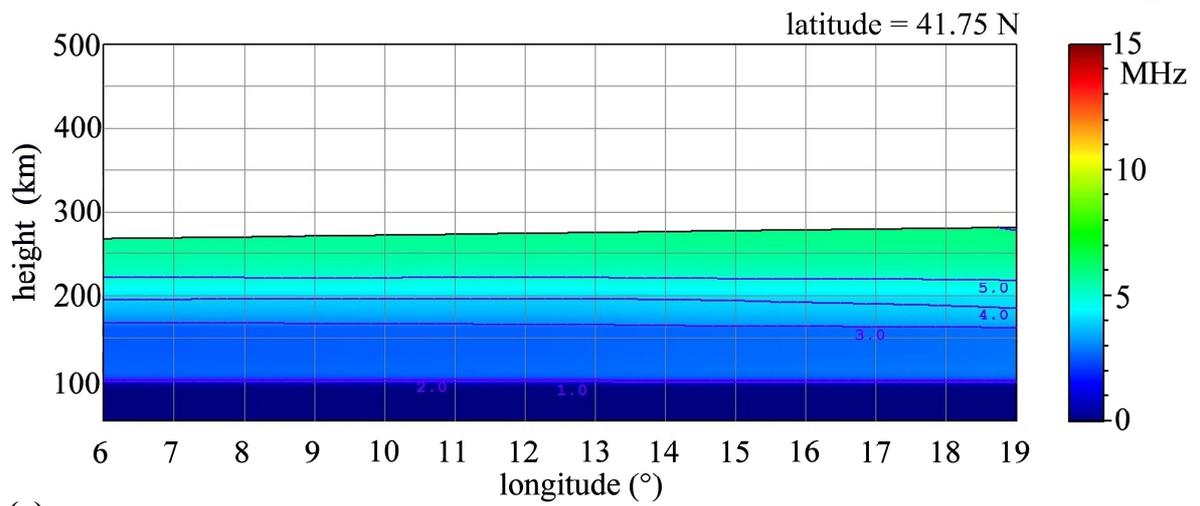

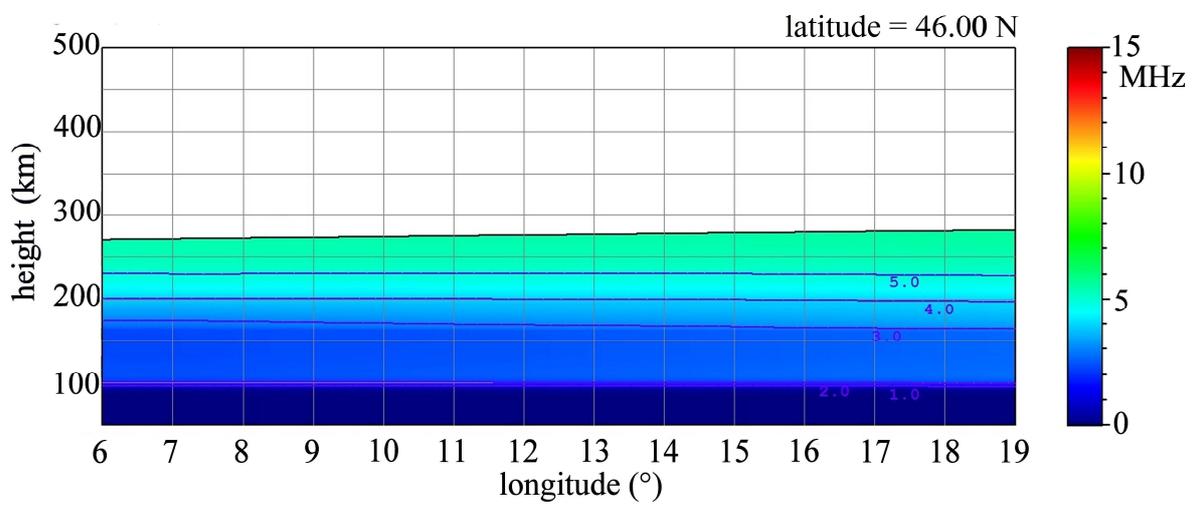

Fig. 16(a)-(c). Cross-sectional maps of $f_p$ at the fixed latitudes 37.5N, 41.75N, and 46N, longitude from 6.0E to 19.0E, and $h$ from 60 km to $h_mF_2$, generated for March 20, 2015 at 09:30 UT by the 3D model studied in this work.

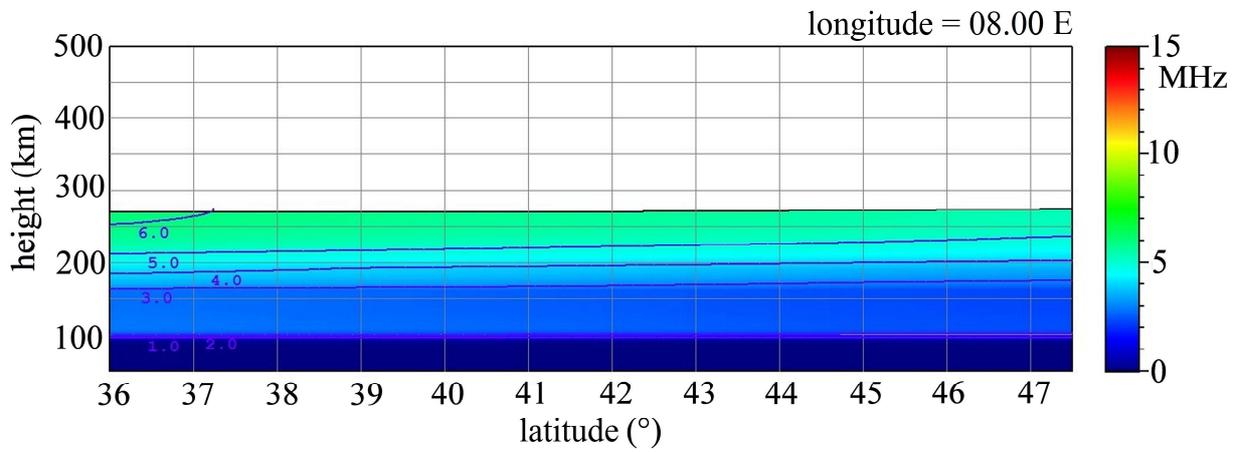

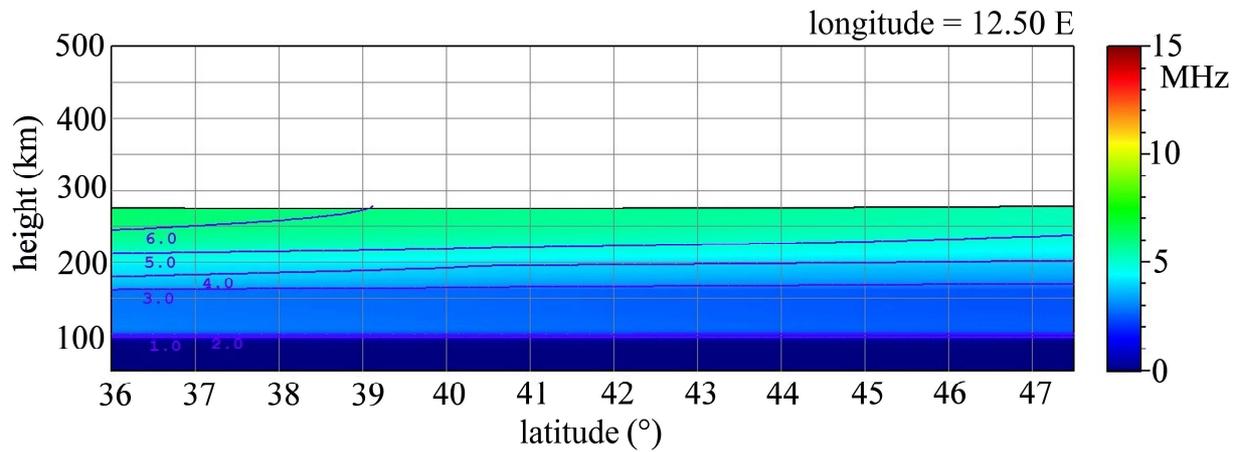

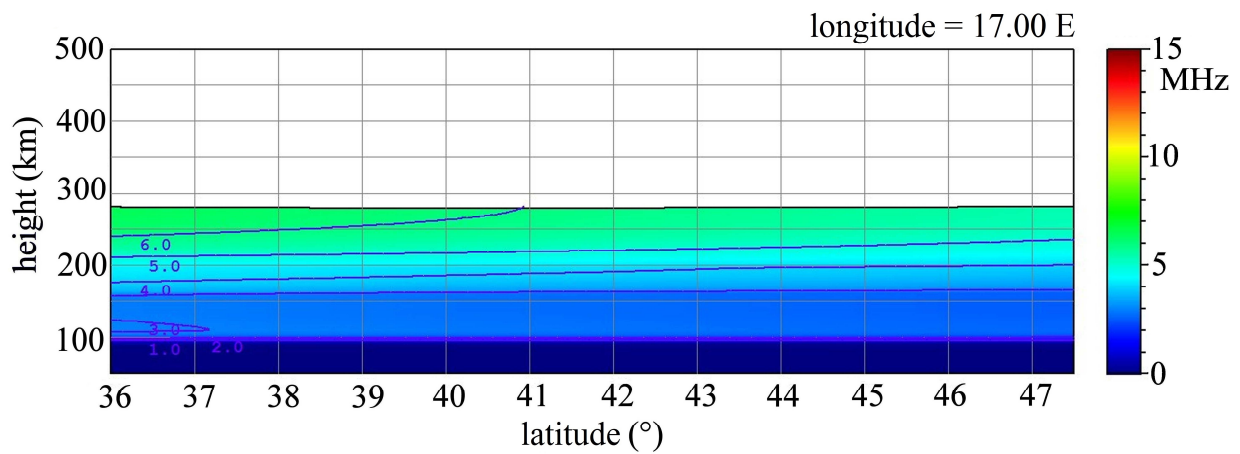

Fig. 17(a)-(c). Cross-sectional maps of $f_p$ at the fixed longitudes 8E, 12.5E, 17E, latitude from 36.0N to 47.5N, and $h$ from 60 km to $h_mF_2$, generated for March 20, 2015 at 09:30 UT by the 3D model studied in this work.